%% file: template.tex
\documentclass{article}
\PassOptionsToPackage{numbers, compress}{natbib}
\usepackage[preprint]{conf}
\usepackage[utf8]{inputenc}
\usepackage[T1]{fontenc}
\usepackage{hyperref}
\usepackage{url}
\usepackage{booktabs}
\usepackage{amsfonts}
\usepackage{nicefrac}
\usepackage{microtype}
\usepackage{xcolor}
\usepackage{wrapfig}
\usepackage{cutwin}
\usepackage{lipsum}
\input{tool}

\DeclareMathOperator*{\True}{True}

\title{%
  \stackon[0pt]{\textnormal{\bf MAGIS}}{\smash{\raisebox{-0.2em}{\hspace{1.86em}\includegraphics[width=1em,height=1em,keepaspectratio]{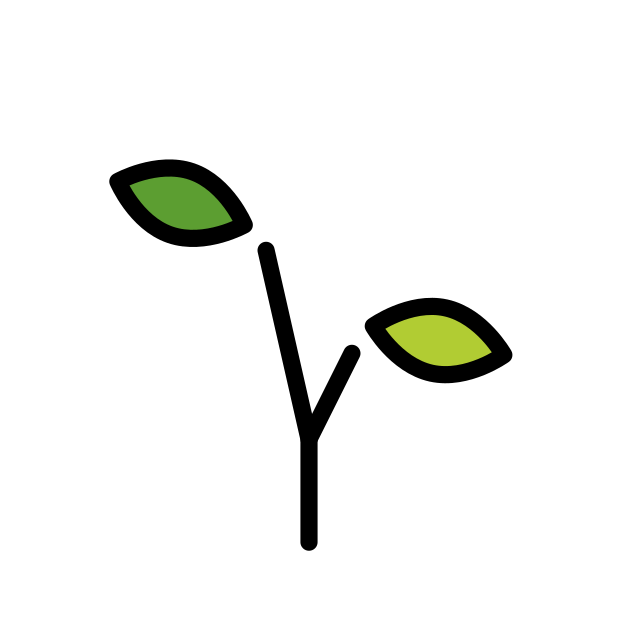}}}}: LLM-Based \underline{M}ulti-\underline{A}gent Framework\\for \underline{G}itHub \underline{I}ssue Re\underline{S}olution
}

\author{%
  Wei Tao \\
  Fudan University \\
  \texttt{wtao18@fudan.edu.cn} \\
  \And
  Yucheng Zhou \\
  University of Macau \\
  \texttt{yucheng.zhou@connect.um.edu.mo} \\
  \And
  Yanlin Wang \\
  Sun Yat-sen University \\
  \texttt{wangylin36@mail.sysu.edu.cn}
  \And
  Wenqiang Zhang \\
  Fudan University \\
  \texttt{wqzhang@fudan.edu.cn} \\
  \And
  Hongyu Zhang \\
  Chongqing University \\
  \texttt{hyzhang@cqu.edu.cn}
  \And
  Yu Cheng \\
  The Chinese University of Hong Kong \\
  \texttt{chengyu@cse.cuhk.edu.hk} 
}

\begin{document}
\maketitle
\thispagestyle{plain}
\begin{abstract}
\input{abstract}
\end{abstract}
\input{main}
\end{document}

%% file: tool.tex
\usepackage{enumitem}
\usepackage{verbatim}
\usepackage{xspace}
\usepackage{diagbox}
\usepackage{array}
\usepackage{cite}

\usepackage{multirow}
\usepackage{diagbox}
\usepackage{url}
\usepackage{comment}
\usepackage{booktabs}
\usepackage{amsmath}
\usepackage{scalerel}
\usepackage[figuresright]{rotating}
\usepackage{threeparttable}
\usepackage{algorithm}
\usepackage{algorithmic}
\usepackage{siunitx}
\usepackage{stackengine}
\usepackage{pifont}

\usepackage{tcolorbox}
\definecolor{grey}{RGB}{128,128,128}

\newcommand{\github}{GitHub\xspace}

\newcommand{\Ourmethod}{MAGIS\xspace}

\newcommand{\Swebench}{SWE-bench\xspace}
\newcommand{\ChatgptThreeFive}{GPT-3.5\xspace}
\newcommand{\ClaudeTwo}{Claude-2\xspace}
\newcommand{\GptFour}{GPT-4\xspace}
\newcommand{\SwellamaSeven}{SWE-Llama 7b\xspace}
\newcommand{\SwellamaThirteen}{SWE-Llama 13b\xspace}

\newcommand{\Fig}{Fig.\xspace}
\newcommand{\Tab}{Tab.\xspace}
\newcommand{\Equ}{Equation\xspace}

\usepackage{mathtools}
\usepackage{multirow}
\usepackage{multicol}
\usepackage{booktabs}
\usepackage{subfigure}

\usepackage{amsmath,amsfonts,bm}

\def\eqref#1{equation~\ref{#1}}

\def\1{\bm{1}}

\def\mL{{\bm{L}}}

\DeclareMathAlphabet{\mathsfit}{\encodingdefault}{\sfdefault}{m}{sl}
\SetMathAlphabet{\mathsfit}{bold}{\encodingdefault}{\sfdefault}{bx}{n}

\def\gC{{\mathcal{C}}}
\def\gD{{\mathcal{D}}}

\def\gG{{\mathcal{G}}}

\def\gI{{\mathcal{I}}}

\def\gL{{\mathcal{L}}}
\def\gM{{\mathcal{M}}}

\def\gP{{\mathcal{P}}}

\def\gR{{\mathcal{R}}}

\def\gT{{\mathcal{T}}}



%% file: abstract.tex
In software development, resolving the emergent issues within GitHub repositories is a complex challenge that involves not only the incorporation of new code but also the maintenance of existing code.
Large Language Models (LLMs) have shown promise in code generation but face difficulties in resolving Github issues, particularly at the repository level. 
To overcome this challenge, we empirically study the reason why LLMs fail to resolve GitHub issues and analyze the major factors. 
Motivated by the empirical findings, we propose a novel LLM-based \textbf{M}ulti-\textbf{A}gent framework for \textbf{G}itHub \textbf{I}ssue re\textbf{S}olution, \textbf{\Ourmethod}, consisting of four agents customized for software evolution: Manager, Repository Custodian, Developer, and Quality Assurance Engineer agents. 
This framework leverages the collaboration of various agents in the planning and coding process to unlock the potential of LLMs to resolve GitHub issues. 
In experiments, we employ the \Swebench benchmark to compare \Ourmethod with popular LLMs, including \ChatgptThreeFive, \GptFour, and \ClaudeTwo. 
\Ourmethod can resolve \textbf{13.94\%} GitHub issues, significantly outperforming the baselines.
Specifically, \Ourmethod achieves an eight-fold increase in resolved ratio over the direct application of \GptFour, the advanced LLM. 

%% file: main.tex
\section{Introduction}\label{sec:intro}

In real-world software development, the code repository for a project is rarely set in stone. High-quality and popular software always evolves to address emergent bugs or new requirements. 
On platforms such as \github\citep{github}, issues typically signify the requirement for software evolution. However, addressing these issues poses significant challenges, as it requires implementing the code change across the entire repository and maintaining the existing functionality while integrating new capabilities. For example, \texttt{django}, a framework for over $1.6$M projects has $34$K issues~\citep{djangoissue}.
Consequently, resolving \github issues remains a significant challenge across academia and industry~\citep{jimenez2024swebench, DBLP:conf/issre/BissyandeLJRKT13}.

Large language models (LLMs) have demonstrated remarkable capabilities across a variety of tasks~\citep{DBLP:journals/corr/abs-2303-12712}, including code generation and code understanding~\citep{DBLP:journals/corr/abs-2308-11396,sun2024survey}. Specifically, LLMs excel in generating function-level code, as evidenced by their performance on numerous benchmark datasets such as MBPP~\citep{MBPP} and HumanEval~\citep{HumanEval}. Despite their success, LLMs remain challenged in tasks that require advanced code generation capabilities, such as class-level code generation~\citep{du2023classeval}. Moreover, LLMs exhibit limitations in processing excessively long context inputs and are subject to constraints regarding their input context length~\citep{abs-2307-03172}. This limitation is particularly evident in repository-level coding tasks, such as solving \github issues, where the context comprises the entire repository, thus imposing constraints on directly using the full repository as input to LLMs.

To harness the full potential of LLMs, many LLM-based multi-agent systems are designed~\citep{hong2023metagpt, qian2023communicative, tufano2024autodev}. These methods have significantly improved LLMs' efficacy in code generation, enabling these systems to construct code repositories based on LLM. 
While these methods address the process of transitioning code repositories from inception to establishment, they rarely consider the handling of software evolution, e.g., resolving \github issues. For \github repositories, especially the popular ones, a large number of commits are pushed every day. These commits derive from a spectrum of evolutionary requirements that span bug fixes, feature additions, performance enhancements, etc~\citep{KADEL24}. For open-source software, new requirements frequently emerge as issues in the project's repository.

Recently, \citet{jimenez2024swebench} developed a benchmark, namely \Swebench, to investigate the capability of popular LLMs in addressing \github issues.
Their study reveals that LLMs fail to resolve over $95\%$ of instances, even when file paths that require modifications are provided. 
This significantly low rate underscores the importance of understanding the reasons behind their suboptimal performance.

In this study, we analyze the factors impacting the effectiveness of LLMs in resolving \github issues. Furthermore, our empirical analysis has concluded a correlation between locating files/lines to be modified and the performance of resolving \github issues. Based on these insights, we propose a novel LLM-based multi-agent framework, termed \Ourmethod, comprising four types of agents: Manager, Repository Custodian, Developer, and Quality Assurance (QA) Engineer. 
Our approach facilitates the resolution of \github issues through collaboration among agents, each fulfilling a unique role: the Manager coordinates the entire process, the Repository Custodian enhances locating files, the Developer performs code changes after locating lines, and the QA Engineer reviews the code change.

In our experiment, we evaluate our framework on \Swebench and compare its performance against existing popular LLMs, such as ChatGPT-3.5 \citep{openai2023gpt3}, GPT-4 \citep{openai2023gpt4}, and Claude-2 \citep{claude2}.
The results demonstrate that our framework, utilizing \GptFour as its base model, significantly outperforms baselines and achieves an eight-fold performance gain compared to the direct application of \GptFour. Further analysis reveals that additional factors, i.e., the planning of code change, locating lines within the code file, and code review process, can significantly influence the resolution rate.

Our main contributions are summarized as follows:
\begin{itemize}[leftmargin=*]
    \item We conduct an empirical analysis of LLMs in resolving \github issues and explore the correlation between locating code file/line, complexity of the code change, and the success rate in resolution.
    \item We propose a novel LLM-based multi-agennt framework, \Ourmethod, to alleviate the limitations of existing LLMs on \github issue resolution. Both our designed four-type agents and their collaboration for planning and coding unlock LLMs' potential on the repository-level coding task.
    \item We compare our framework and other strong LLM competitors (i.e., \ChatgptThreeFive, \GptFour, and \ClaudeTwo) on the \Swebench dataset. The results show \Ourmethod significantly outperforms these competitors. Further analysis confirms the effectiveness and necessity of our framework design.
\end{itemize}

\section{Empirical Study}\label{sec:rq1}

\Swebench~\citep{jimenez2024swebench} reveals the challenges LLMs face in addressing \github issue resolution. For example, in their evaluation, \GptFour can only resolve less than 2\% issues of the test set. 
Conversely, in tasks like function-level code generation, LLMs exhibit superior performance (e.g., \GptFour gets the score of $67.0$ on HumanEval~\citep{gpt4report}).
Given the complexity of \github issue resolution akin to repository-level coding,  we aim to investigate \textbf{Why the Performance of Directly Using LLMs to Resolve \github Issue is Limited? (RQ 1)}. We answer this RQ from the following three aspects:

\paragraph{Locating the Files to be Modified.}

\github issue resolution is a repository-level coding task, distinguishing it from file-level coding tasks primarily in the challenge of locating the files requiring modification.
\citet{jimenez2024swebench} employ the BM25 method~\citep{DBLP:conf/trec/RobertsonWJHG94} to retrieve relevant code files that are subsequently utilized as input to the LLM.
After employing retrieval methods, it is necessary to select the top-$K$ files or truncate the content based on the maximum context length of the LLM. Incorporating more files can enhance recall scores. However, it also imposes significant demands on the capabilities of LLMs. As demonstrated by the study~\citep{jimenez2024swebench}, \ClaudeTwo exhibits a decrease in the resolved ratio (from 1.96 to 1.22) as recall scores increase (from 29.58 to 51.06). This decline may be attributed to the inclusion of irrelevant files or the limited capacity of LLMs to process longer contexts effectively. Consequently, optimizing the performance of LLMs can be better achieved by striving for higher recall scores with a minimized set of files, thus suggesting a strategic balance between recall optimization and the number of chosen files.

\paragraph{Locating the Lines to be Modified.}

Beyond the impact of file locating, we delve into the generation of failed instances when the correct modified files were provided.
A typical code change consists of multiple hunks, each specifying the line numbers targeted for modification and detailing the changes made at these locations.
To quantitatively analyze the accuracy of line localization, we use the line numbers' range of the modified content in the reference code change as the basis assuming that the correct modification location of the code change is uniquely determined in most cases. By calculating the coverage ratio of the line number ranges of the generated and reference, we can estimate the accuracy of line localization in the generation process, i.e.,
\begin{align}
    \text{Coverage Ratio} = \frac{\sum_{i=0}^{n} \sum_{j=0}^{m} \left| [s_i,e_i] \cap [s_j',e_j'] \right|}{\sum_{i=0}^{n} (e_i - s_i + 1)}, ~\label{eq:overlap}
\end{align}
where the numerator is the length of the intersection of modified lines between the reference divided into $n$ hunks and the generation divided into $m$ hunks, and the denominator is the number of modified lines in the reference.
More details about \Equ\ref{eq:overlap} can be found in Appendix \ref{appendix:overlap_ratio}.

\begin{wrapfigure}{r}{0.44\textwidth}
    \centering
    \includegraphics[width=\linewidth]{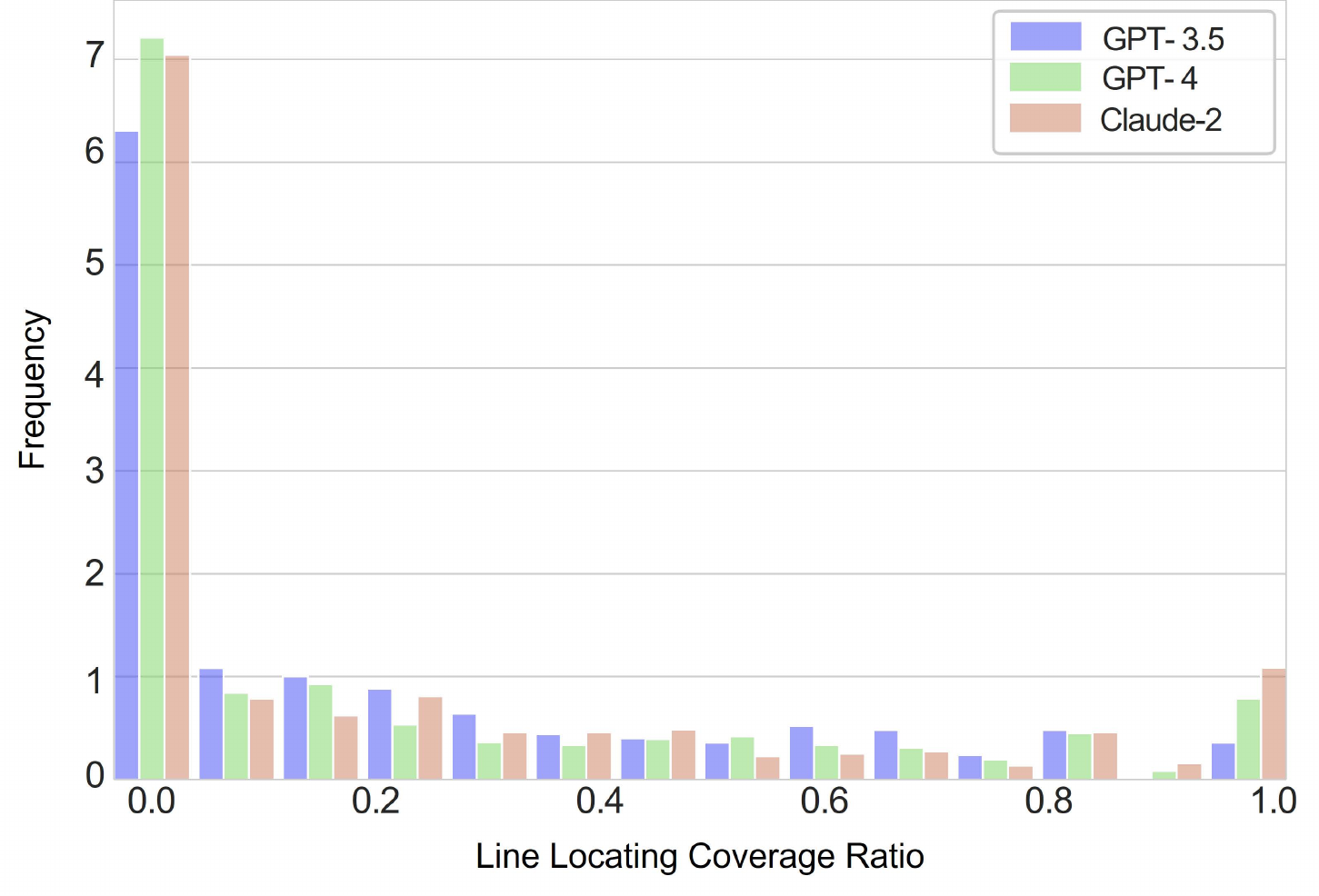}
    \vspace{-2em}
    \caption{\small The comparison of line locating coverage ratio between three LLMs. The 
vertical axis representing the frequency of the range of line locating coverage ratio for each group, and the horizontal axis representing the coverage ratio.}
    \label{fig:line_location_overlap_baselines}
    \vspace{-1em}
\end{wrapfigure}
For $574$ instances in the \Swebench that experiments \GptFour \citep{jimenez2024swebench}, the distribution of the coverage ratio between the results generated by three LLMs and the reference is shown in \Fig~\ref{fig:line_location_overlap_baselines}.
From this, we observe that the performance of LLMs in generating the code change is probably related to their ability to locate code lines accurately (Detailed explanation can be found in Appendix \ref{appendix:observation_line_location}).

Furthermore, we assess the relationship between the coverage ratio and the issue resolution by calculating their correlation coefficient. Given that the distribution of these variables exhibits skewness, and the resolution result is binary (resolved or not), logistic regression is employed for the analysis across three LLMs.
However, due to the limited number of successfully generated instances on \GptFour and \ChatgptThreeFive, a statistically significant relationship is only detected in the result generated by \ClaudeTwo. The result, i.e., P-value < $0.05$, shows statistical significance. 
Specifically, with a coefficient, $0.5997$, on \ClaudeTwo, there is a substantial and positive relation between improvements in the coverage ratio and the probability of successfully resolving issues, which demonstrates that locating lines is a key factor for \github issue resolution.

\paragraph{Complexity of the Code Changes.}
The complexity of the code change is reflected in various indices: the number of modified files, functions, hunks, and lines added or deleted.
Firstly, we quantitatively assess the complexity by calculating the value of various indices corresponding to the reference code change. Secondly, the coefficient is calculated between the numbers in each index and the issue resolution. 
\Tab~\ref{tab:empirical_complexity_relation} shows the correlation scores under the logistic regression.

\begin{table}[ht]\small
    \centering
    \caption{\small Correlation between the complexity indices and the issue resolution.}
    \label{tab:empirical_complexity_relation}
    \begin{threeparttable}
    \begin{tabular}{lcccccc}
    \toprule  
    LLM     & \# Files & \# Functions  & \# Hunks  & \# Added LoC    & \# Deleted LoC  & \# Changed LoC \\
    \midrule
    \ChatgptThreeFive	   & $-17.57$\tnote{*}	& $-17.57$\tnote{*}	&  $-0.06$\tnote{*}	&  $-0.02$\tnote{~}	&  $-0.03$\tnote{~}	&  $-0.53$\tnote{*}	\\
    \GptFour	           & $-25.15$\tnote{*}	& $-25.15$\tnote{*}	&  $-0.06$\tnote{~}	&  $-0.10$\tnote{~}	&  $-0.04$\tnote{~}	&  $-0.21$\tnote{~}	\\
    \ClaudeTwo	           &  \hspace{0.5em}$-1.47$\tnote{*}	&  \hspace{0.5em}$-1.47$\tnote{*}	&  $-0.11$\tnote{*}	&  $-0.09$\tnote{*}	&  $-0.07$\tnote{*}	&  $-0.44$\tnote{*}	\\
    \bottomrule
    \end{tabular}
    \begin{tablenotes}
    \item[*] The correlation between the index and the issue resolution is significant (P-value \(<\) $0.05$).
    \end{tablenotes}
    \end{threeparttable}
\end{table}

As shown in \Tab~\ref{tab:empirical_complexity_relation}, all three LLMs demonstrate a statistically significant correlation with the issue resolution across several indices.
The correlation scores for the number of files and functions modified are notably negative for all models, indicating that an increase in these indices is associated with a decreasing likelihood of issue resolution. This suggests that the more complex the code change, as indicated by a higher number of files and functions modified, may hinder the issue resolution. 
More analysis can be found in Appendix \ref{appendix:analysis_complexity}.
The analysis reveals a relationship between the complexity, as measured by several indices, and whether to successfully resolve the issues in software evolution. The negative correlations suggest that increased complexity, particularly in terms of the number of files and functions changed, tends to hinder issue resolution.

\section{Methodology}\label{sec:method}
\begin{figure*}
    \centering
    \includegraphics[width=\textwidth]{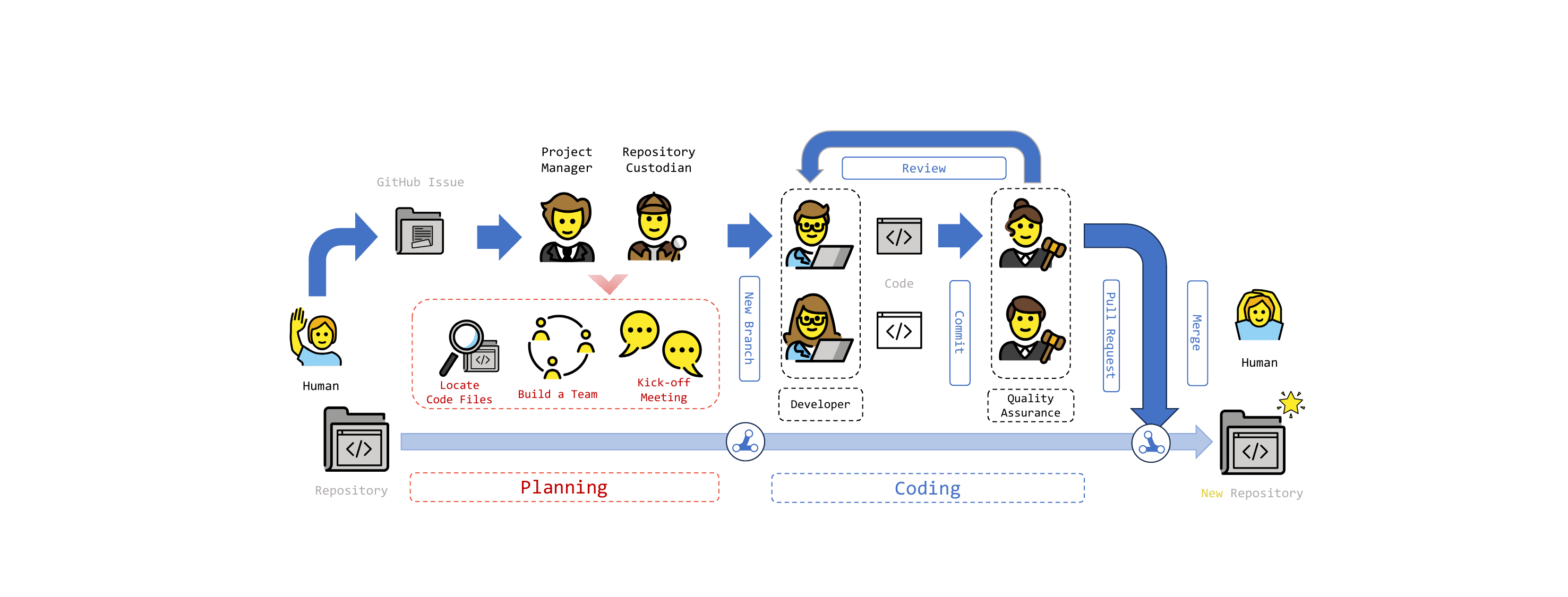}
    \vspace{-1em}
    \caption{\small Overview of our framework, \Ourmethod. The detailed version can be found in \Fig\ref{fig:detailed_overview}.}
    \label{fig:overview}
    \vspace{-1em}
\end{figure*}

Based on the empirical study identifying key factors affecting LLMs' issue resolution, we design the framework illustrated in \Fig~\ref{fig:overview}. This framework aims to mitigate negative impacts by transforming the complex task of \github issue resolution into a collaborative effort. It incorporates four key roles for LLM-based agents working collaboratively in the workflow:
\ding{172} \textit{Manager}: this role tasks with team assembly, meeting organization, and plan formulation.
\ding{173} \textit{Repository Custodian}: it is responsible for locating the relevant files in the repository acording to the \github issue and recording the change of the repository.
\ding{174} \textit{Developer}: this role participates in planning discussions and completes tasks from the Manager.
\ding{175} \textit{Quality Assurance (QA) Engineer}: it reviews the code change from Developers to ensure the quality of the whole repository.

The collaborative process involves planning and coding. In the planning, an issue is assigned to the Manager and the Repository Custodian. The custodian identifies candidate files relevant to the issue for modification. With the issue description and a list of candidate files, the Manager defines tasks and assembles a team, where each member is a Developer specifically designed for the defined task. The Manager holds a kick-off meeting with Developers and devises a plan. During coding, Developers undertake their assigned tasks from the Manager, and the QA Engineer reviews each code change. If a change fails to meet quality standards, the QA Engineer provides feedback, prompting further revisions until the QA Engineer approves or a set iteration limit is reached.

\subsection{Agent Role Design}\label{sec:role}

Our workflow draws inspiration from the \github Flow\citep{githubflow}, an effective human workflow paradigm adopted by many software teams. 
Both the human workflow and our LLM-based agent framework prioritize collaboration among individuals with diverse skills. While the underlying principles are similar, there are notable differences.
Accordingly, we have tailored the roles as follows:

\begin{itemize}[leftmargin=*]
    \item \raisebox{-.4\baselineskip}{\includegraphics[height=1.3\baselineskip]{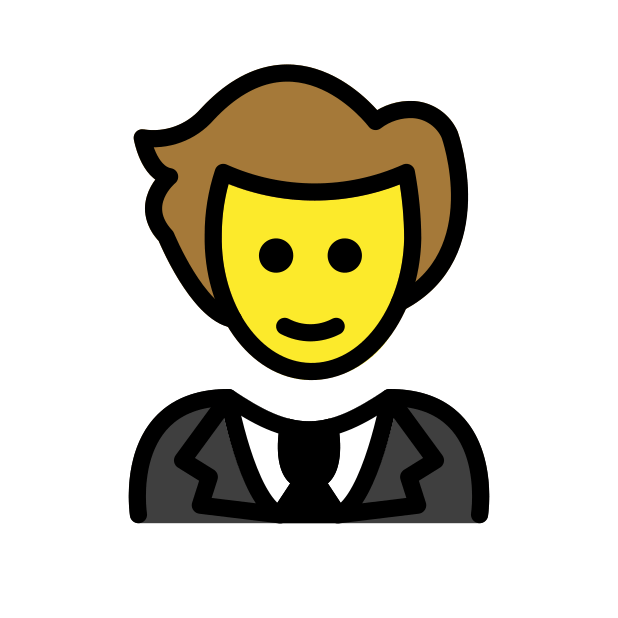}}~\textbf{Manager}. 
    The Manager's role is pivotal in planning. In conventional setups, managers decompose the issue into tasks according to the pre-formed team and allocate these tasks for members with different skills. In contrast, our Manager agent can first decompose the issue into tasks and then design Developer agents to form a team. This setup improves team flexibility and adaptability, enabling the formation of teams that can meet various issues efficiently.
    
    \item \raisebox{-.4\baselineskip}{\includegraphics[height=1.3\baselineskip]{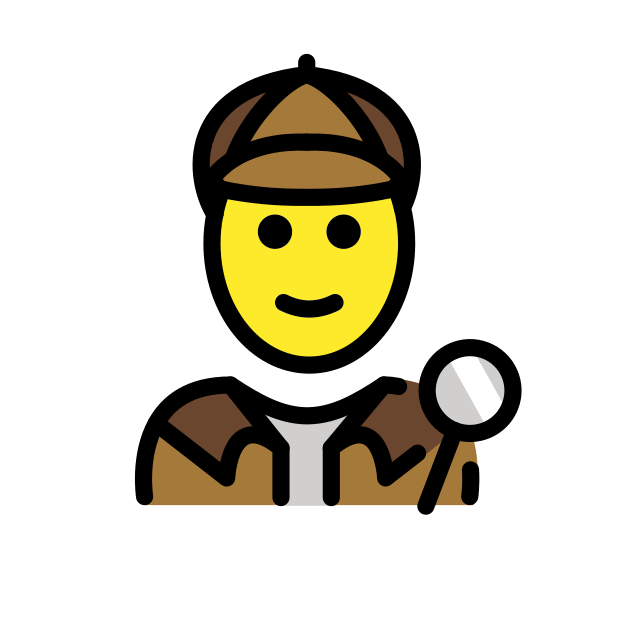}}~\textbf{Repository Custodian}. 
    Considering extensive files in a repository, the custodian agent's task is to locate files relevant to the issue. Unlike humans, who can browse through the entire repository, the LLM-based agent faces challenges in browsing. Although LLMs have extended context limits, their application is constrained in two aspects. First, it is a high computational cost to query each file in an entire repository for each update, particularly when some repositories update frequently. Second, the performance of LLMs degrades when the context input is long~\citep{DBLP:journals/corr/abs-2404-02060, abs-2307-03172, DBLP:journals/corr/abs-2311-08734}.
    
    \item \raisebox{-.4\baselineskip}{\includegraphics[height=1.3\baselineskip]{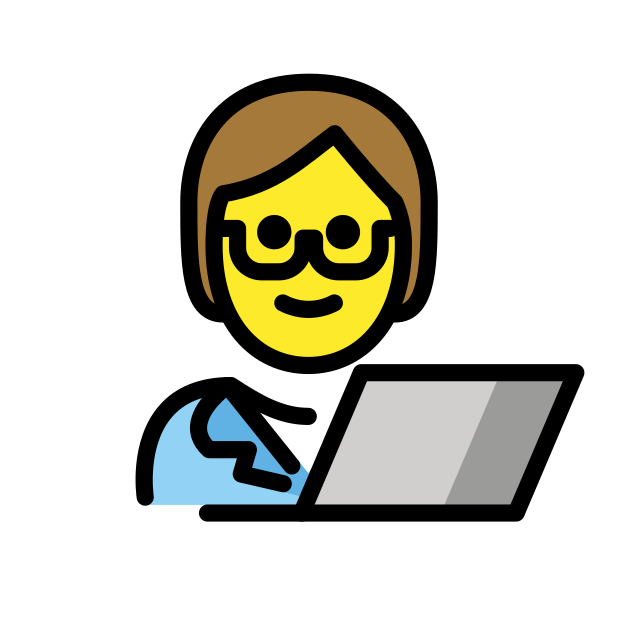}}~\textbf{Developer}. 
    Compared to human developers, the Developer agent can work continuously and efficiently. Therefore, scheduling the agent to work in parallel is easier than scheduling humans who require considering factors beyond the task.
    Additionally, although numerous developer agents are capable of generating code~\citep{hong2023metagpt, qian2023communicative}, their ability to modify existing code is not equally proficient. To address this issue, our framework decomposes the code modification process into sub-operations including code generation. This approach enables Developers to leverage the benefits of automatic code generation thereby producing applicable code changes.
    
    \item \raisebox{-.4\baselineskip}{\includegraphics[height=1.3\baselineskip]{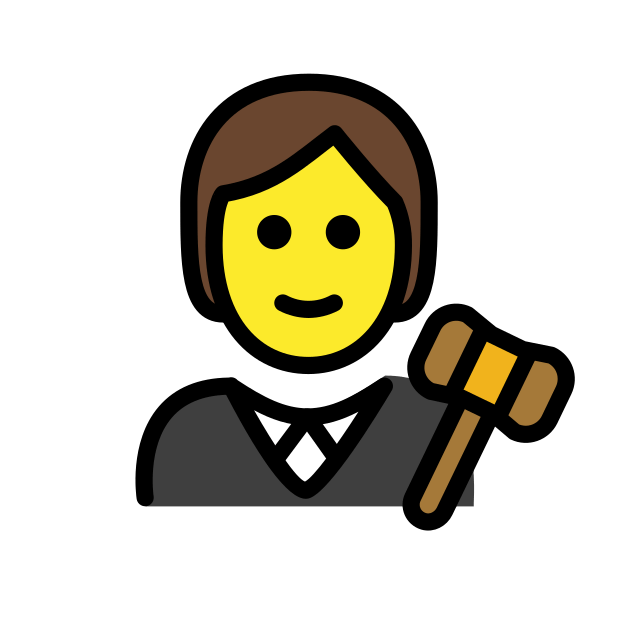}}~\textbf{QA Engineer}. 
    In software evolution, QA Engineers play a crucial role in maintaining software quality through code review~\citep{DBLP:conf/msr/McIntoshKAH14, DBLP:conf/icsm/KononenkoBGCG15}. Despite their importance, code review practices are often undervalued or even overlooked~\citep{DBLP:conf/sigsoft/BaumLNS16}. Such neglect can hinder software development, illustrated by instances where developers may experience delays of up to 96 hours awaiting code review feedback~\citep{DBLP:conf/esem/BosuC14}. To address this problem, our framework pairs each Developer agent with a QA Engineer agent, designed to offer task-specific, timely feedback. This personalized QA approach aims to boost the review process thereby better ensuring the software quality.
\end{itemize}

\subsection{Collaborative Process}\label{sec:process}

\subsubsection{Planning}

Three types of role agents engage in the planning: Repository Custodian, Manager, and Developer. 
This process comprises three phases: locating code files, team building, and kick-off meeting.

\begin{wrapfigure}{r}{0.395\textwidth}
\begin{minipage}{\linewidth}
\vspace{-3.0em}
\begin{algorithm}[H]
    \footnotesize
    \caption{Locating.}
    \begin{algorithmic}[1]
    \STATE \textbf{Input:} repository: $\gR_{i}$ including files $\{f_{i}\}$, \github issue: $q_x$, LLM: $\gL$
    \STATE \textbf{Config:} filter top width: $k$, prompts: $\gP$, find the latest previous version of the file and its summary: $find$
    \STATE \textbf{Output:} candidate files: $\gC_i^k$ $\gets$ $\emptyset$, repository evolution memory: $\gM$ $\gets$ $\emptyset$
    
    \STATE{$\gR_i$ $\gets$ BM25($\gR_i$, $q_x$)} 
    
    \STATE{$\gC_i^k$ $\gets$ $\gR_i$[:$k$]}
    
    \FOR{$f_i$ $\in$ $\gC_i^k$}
    
        \STATE{$f_h, s_h$ $\gets$ $find$ $(f_i, \gM)$} 
        \IF{$\exists$ $f_h$ $\AND$ len($s_h$) $<$ len$(f_i)$}
            \IF{$h$ is $i$}
                \STATE{$s_i$ $\gets$ $s_h$}
            \ELSE
                \STATE{$\Delta d$ $\gets$ \text{diff}($f_h$, $f_i$)}
                \STATE{$m$ $\gets$ $\gL(\Delta d, \gP_{1})$}
                \STATE{$s_i$ $\gets$ $s_h \cup m$}
            \ENDIF
        \ELSE
            \STATE{$s_i$ $\gets$ $\gL(f_i, \gP_{2})$}
        \ENDIF
    
        \STATE{$\gM$ $\gets$ $\gM$.update($\{f_i: s_i\}$)}
    
        \IF{ $\gL((s_i, q_x), \gP_{3})$ is $\FALSE$}
            \STATE{$\gC_i^k$ $\gets$ $\gC_i^k$ - $f_i$}
        \ENDIF
    
    \ENDFOR
    \end{algorithmic}
    \label{alg:locating}
\end{algorithm}
\vspace{-0.8cm}
\end{minipage}
\end{wrapfigure}

\paragraph{Locating Code Files.}
Firstly, the Repository Custodian employs the BM25 algorithm~\citep{DBLP:conf/trec/RobertsonWJHG94} to rank the files in the repository based on the \github issue description. Subsequently, the top $k$ files are selected as potential candidates for further coding. 
However, as described in \S\ref{sec:rq1}, this simple retrieval method can introduce irrelevant files, increasing the cost and reducing the effectiveness of subsequent coding process. Therefore, we filter these files based on relevance to minimize their number.
While it is feasible to directly assess the relevance between each file and the issue by LLMs, queries to the LLM may contain the same code snippets as previous ones, leading to unnecessary computational costs. Considering that applying the code change often modifies a specific part of the file rather than the entire file, we propose a memory mechanism to reuse the previously queried information.

Algorithm~\ref{alg:locating} outlines the process of locating files with our designed memory $\gM$. 
If a file $f_i$ is compared for the first time with an issue $q_x$, the LLM $\gL$ with prompt $\gP_{2}$ compresses it into the summary $s_i$, where $i$ denotes the file's version.
This summary is shorter than the code content in the file and it is stored in memory for future reuse.
If the file $f_i$ has been previously compared, the latest previous version ($h$) of the file $f_h$ can be found by the script $find$.
Since $f_i$ can be represented as the combination of $f_h$ and the difference between them ($\Delta d$ that be obtained via the ``\texttt{git diff}'' command), LLMs can understand $f_i$ by using $f_h$ and $\Delta d$. If the difference is small and the file $f_i$ is long, it is valuable to reuse the previous summary $s_h$ stored in memory rather than the content of $f_i$.
Specifically, if the length of $s_h$ is less than that of $f_i$, $\gL$ with prompt $\gP_{1}$ can summarize the code changes $\Delta d$ as a ``commit message'' $m$. The combination of $s_h$ and $m$ forms the description of the newer version $f_i$, enabling the LLM $\gL$ with prompt $\gP_{3}$ to determine whether it is relevant to the issue in fewer context length.
Based on their relevance, the custodian agent filters irrelevant files, allowing the Manager agent to define tasks with remaining relevant files.

\paragraph{Team Building.}
In this process, the Manager agent has the flexibility to ``recruit'' team members as the issue needs.
Firstly, upon receiving the located files, the Manager begins with analyzing the \github issue for the repository and breaks them into detailed file-level tasks. 
Specifically, for each code file $f_i$ in the candidate set $\gC_i^k$, the Manager leverages the LLM $\gL$ with the prompt $\gP_{4}$ and the issue description $q_x$ to define the corresponding file-level task $t_i$.
One issue can be converted to multiple tasks. 
These tasks, along with the associated code file, are stored in a task set $\gT_i^k$. 
\begin{wrapfigure}{r}{0.455\textwidth}
\begin{minipage}[t]{\linewidth}
\vspace{-1.6em}
\begin{algorithm}[H]
    \footnotesize
    \caption{Making the plan.}
    \begin{algorithmic}[1]
    \STATE \textbf{Input:} candidate files: $\gC_i^k$, issue: $q_x$, LLM: $\gL$
    \STATE \textbf{Config:} prompts: $\gP$
    \STATE \textbf{Output:} tasks: $\gT_i^k$ $\gets$ $\emptyset$, Developer agents' role description: $\gD_i^k$ $\gets$ $\emptyset$, plan: $c_{main}$

    \FOR{$f_i$ $\in$ $\gC_i^k$}
    
        \STATE{$t_i$ $\gets$ $\gL$(($f_i$, $q_x$), $\gP_{4}$)}

        \STATE{$\gT_i^k$ $\gets$ $\gT_i^k$ $\cup$ ({$f_i$, $t_i$})}

        \STATE{$r_i$ $\gets$ $\gL$(($t$, $q_x$), $\gP_{5}$)}

        \STATE{$\gD_i^k$ $\gets$ $\gD_i^k$ $\cup$ $r_i$}
    
    \ENDFOR

    \STATE{\textit{recording} = {kick\_off\_meeting}({$\gD_i^k$})}

    \STATE{$\gD_i^k$ $\gets$ $\gL$(($\gD_i^k$, \textit{recording}), $\gP_{6})$}

    \STATE{$c_{main}$ $\gets$ $\gL($\textit{recording}, $\gP_{7})$}

    \end{algorithmic}
    \label{alg:planning}
\end{algorithm}
\vspace{-3em}
\end{minipage}
\end{wrapfigure}
Once a task is clarified, the Manager defines the personality role $r_i$ of the Developer by invoking LLM $\gL$ with the prompt $\gP_{5}$ and the task $t_i$. 
By iterating through these candidate code files, the Manager agent ultimately designs a collection of Developer agent role descriptions $\gD_i^k$, thus forming the development team. The details of the team building are shown in Algorithm~\ref{alg:planning}.
This approach simplifies the task for LLMs because each team member only needs to handle a sub-task rather than resolving the entire complex issue.

\paragraph{Kick-off Meeting.}
After building the team, the Manager organizes a kick-off meeting.
This meeting serves two purposes: \ding{172} To confirm whether the tasks assigned by the Manager are reasonable and ensure that all Developers in the team can collaboratively resolve the issue $q_x$, \ding{173} To determine which Developers' tasks can be executed concurrently and which tasks have dependencies need to be sorted. The meeting takes the form of a circular speech: the Manager is responsible for opening the speech, guiding the discussion and summarizing the results, and the Developers provide their opinions based on previous discussions in turn.
One example of the meeting can be found in Appendix~\ref{appendix:meeting}.
After the meeting, Developers adjust their role descriptions $\gD_i^k$ based on the discussion $recording$, and the Manager, leveraging the LLM $\gL$ and the prompt $\gP_{7}$, generates a main work plan $c_{main}$. This plan is presented as code, and embedded into the program for execution.
The meeting makes collaboration among Developers more efficient and avoids potential conflicts.

\subsubsection{Coding}
\begin{wrapfigure}{r}{0.51\textwidth}
\begin{minipage}{\linewidth}
\vspace{-5.7em}
\begin{algorithm}[H]
    \footnotesize
    \caption{Coding task execution.}
    \begin{algorithmic}[1]
    \STATE \textbf{Input:} file-task pairs set: $\gT_i^k$, LLM: $\gL$
    \STATE \textbf{Config:} prompts: $\gP$, the max of iteration: $n_{\text{max}}$
    \STATE \textbf{Output:} code changes: $\gD$

    \FOR{$f_i$, $t_i$ $\in$ $\gT_i^k$}

        \STATE{$a_i$ $\gets$ $\gL$(($f_i$, $t_i$), $\gP_{8}$)}
        
        \FOR{$j$ $\in$ [ $0$, $n_{\text{max}}$ )}
            \IF{$j$ $>$ $0$}
                \STATE{$t_i$ = ($t_i$, \textit{review\_comment})}
            \ENDIF
            
            \STATE{$\{[s_i', e_i']\}$ $\gets$ $\gL$(($f_i$, $t_i$), $\gP_{9}$)}
            \STATE{$f_i$, \textit{old\_part} $\gets$ split($f_i$, $\{[s_i', e_i']\}$)}
            \STATE{\textit{new\_part} $\gets$ $\gL$(($f_i$, $t_i$, \textit{old\_part}), $\gP_{10}$)}
            \STATE{$f_i'$ $\gets$ replace($f_i$, $\{[s_i', e_i']\}$, \textit{new\_part})}
            
            \STATE{$\Delta d_i$ $\gets$ \text{diff}($f_i$, $f_i'$)}

            \STATE{\textit{review\_comment} = $\gL$(($t_i$, $\Delta d_i$), $\gP_{11}$)}

            \STATE{\textit{review\_decision} = $\gL$((\textit{review\_comment}), $\gP_{11}$)}
            \IF{\textit{review\_decision} is $\TRUE$}
                \STATE{\textbf{break}}
            \ENDIF
    
        \ENDFOR
        \STATE{$\Delta d$ $\gets$ \text{diff}($f_i'$, $f_i$)}
        \STATE{$\gD$ $\gets$ $\gD \cup \Delta d$}
    
    \ENDFOR

    \end{algorithmic}
    \label{alg:execution}
\end{algorithm}
\vspace{-2.7em}
\end{minipage}
\end{wrapfigure}
Based on the empirical study on line locating and the complexity (\S\ref{sec:rq1}), we transform the code change generation into the multi-step coding process that is designed to leverage the strengths of LLMs in code generation while mitigating their weaknesses in code change generation.
Two types of agents participate in the coding process: Developers and QA Engineers. 
As outlined in Algorithm~\ref{alg:execution}, for each task $t_i$ and its associated code file $f_i$ in $\gT_i^k$, the Developer agent generates the role description of the QA Engineer $a_i$ by the LLM $\gL$ with the prompt $\gP_{8}$. 
Subsequently, Developers collaborate with their QA Engineers to execute the coding tasks.
During each execution of the Developer, the range of lines of code that need to be modified is firstly determined as a set of intervals $\{[s_i', e_i']\}$ where $s_i'$ represents the starting line number in the $i$-th hunk, and $e_i'$ is the ending line number.
The determination is generated by analyzing the task content $t_i$ and file content $f_i$ using $\gL$ with the prompt $\gP_{9}$.
These intervals split the original code file $f_i$ into parts to be modified (\textit{old\_part}) and parts to be retained. Developers then generate new code snippets, \textit{new\_part}, by $\gL$ with the prompt $\gP_{10}$. The code snippets replace \textit{old\_part}, resulting in a new version of the code file $f_i'$. Utilizing Git tools, the code change $\Delta d_i$ for this file $f_i$ is generated. 
With the code change $\Delta d_i$, QA Engineer produce \textit{review\_comment} and \textit{review\_decision}, by the LLM $\gL$ with the prompt $\gP_{11}$. 
If the decision, \textit{review\_decision}, is negative (i.e., $false$), the feedback, \textit{review\_comment}, prompts Developers to revise the code in the next attempt. 
This iterative process continues until the code change meets the quality standards (i.e., \textit{review\_decision} is $true$) or reaches a predefined maximum number of iterations. 
After the iteration, the final version of the code change, $\Delta d$, is fixed, which is the ultimate modification result on each file. 
All generated final-version code changes during this process are merged into the repository-level code change $\gD$ as the issue solution.

\section{Experiments and Analysis}\label{sec:exp_analysis}

\subsection{Setup}\label{sec:setup}

In the experiments, we employ the \Swebench dataset as the evaluation benchmark because it is the latest dataset specifically designed for evaluating the performance of the \github issue resolution. \Swebench comprises $2,294$ issues extracted from $12$ popular Python repositories, representing real software evolution requirements. 
Given the observation that experimental outcomes on the $25\%$ subset of \Swebench align with those obtained from the entire dataset~\citep{jimenez2024swebench}, we opt for the same $25\%$ subset previously utilized in experiments for \GptFour according to their materials~\citep{swesub}. Moreover, the experimental scores for the five LLMs, have been made available by them~\citep{swebenchopen}.

Our framework is flexible to integrate various LLMs. To compare with the scores reported by \Swebench, \GptFour is selected as the base LLM. Another reason for the selection is that \GptFour shows remarkable performance on code generation and understanding as demonstrated on benchmarks such as MBPP~\citep{MBPP} and HumanEval~\citep{HumanEval}. \ClaudeTwo is not chosen due to the unavailability of API access.

Following \Swebench~\citep{jimenez2024swebench}, the applied and resolved ratio is used to evaluate the performance under the setting with the files requiring modification provided. 
The applied ratio indicates the proportion of instances where the code change is successfully generated and can be applied to the code repository by Git. The resolved ratio refers to the proportion of instances where the code change is successfully applied and passes a series of tests. Additional elaboration is provided in Appendix \ref{appendix:metric}.

\subsection{How Effective is Our Framework? (RQ 2)}\label{sec:rq2}

The comparative performance analysis between our framework and other LLMs on the same dataset is presented in \Tab~\ref{tab:swe_result_574}. The results indicate that our framework significantly outperforms other LLMs. Notably, with a resolved ratio of $13.94\%$, our framework's effectiveness is eight-fold that of the base LLM, \GptFour. This substantial increase underscores our framework's capability to harness the potential of LLMs more effectively.
Furthermore, when contrasted with the previous state-of-the-art LLM, \ClaudeTwo, our framework's resolved ratio exceeds that benchmark by more than two-fold. This superior performance unequivocally establishes the advance of our method.

\begin{wraptable}{r}{0.57\textwidth}\small
\vspace{-0.6cm}
    \centering
    \caption{\small The comparison of overall performance between \Ourmethod and baselines on \Swebench.}
    \label{tab:swe_result_574}
    \vspace{0.2cm}
    \begin{threeparttable}
    \begin{tabular}{lcc}
    \toprule
    \textbf{Method}     & \textbf{\% Applied}   & \textbf{\% Resolved} \\
    \midrule
    \ChatgptThreeFive   & $11.67$   & $0.84$    \\
    \ClaudeTwo          & $49.36$   & $4.88$    \\
    \GptFour            & $13.24$   & $1.74$    \\
    \SwellamaSeven      & $51.56$   & $2.12$    \\
    \SwellamaThirteen   & $49.13$   & $4.36$    \\
    \midrule
    \bf \Ourmethod      & \textbf{97.39}   & \textbf{13.94}  \\
    \Ourmethod (w/o QA)      & $92.71$   & $10.63$  \\ 
    \Ourmethod (w/o hints)      & $94.25$   & $10.28$  \\ 
    \Ourmethod (w/o hints, w/o QA) & $91.99$   & $8.71$  \\ 
    \toprule
    \end{tabular}
    \end{threeparttable}
\vspace{-0.2cm}
\end{wraptable}

The ablation study is designed to simulate two scenarios: 
\ding{172} Without QA (w/o QA): Considering the QA Engineer agent as optional within our framework, we directly evaluate the code changes generated by the Developer agent, bypassing the QA process. This scenario aims to investigate the effectiveness and necessity of QA Engineer review.
\ding{173} Without hints (w/o hints): Hints refer to the textual content found in the comments section of pull requests, which are typically created before the first commit of the pull request. This setting means our framework operates without any clarifications except for the issue, despite such information being available on \github before the issue resolution process begins. This analysis aims to explore if the participation of humans could potentially improve the success rate of issue resolution.

Our framework shows a significant improvement in issue resolution, even without QA or hints. It achieves a resolved ratio of $8.71$, which is five times higher than that of the base LLM. This increase underscores the contribution of other agents in \Ourmethod to its overall performance. Furthermore, integrating cooperation with QA or hints separately can further elevate the resolved ratio by $1.92$ or $1.57$, respectively. These findings underscore the value of QA Engineers and the participation of humans, as demonstrated by the resolved rates achieved through their integration.

For instance, to resolve the issue~\citep{django30255} from the repository \texttt{Django}~\citep{django}, the developer modifies four hunks in two files~\citep{django12155}, as shown in \Fig~\ref{fig:django_case_gold}. 
Despite the availability of two provided files, our method opts for modifications in only one file, as illustrated in Figure~\ref{fig:django_case_our}. Remarkably, this simpler code change enables the repository to pass all requisite test cases.

Additional comparison can be found in Appendix~\ref{appendix:exp_lite} and \ref{appendix:devin}, and detailed case study is shown in Appendix~\ref{appendix:case_study}. Furthermore, the statistics on the generated code changes can be found in Appendix~\ref{appendix:stat}.

\subsection{How Effective is Our Planning Process? (RQ 3)}\label{sec:rq3}
\begin{figure}[t]
    \centering
    \begin{minipage}[b]{0.47\textwidth}
        \vspace{-2.5em}
        \includegraphics[width=\linewidth]{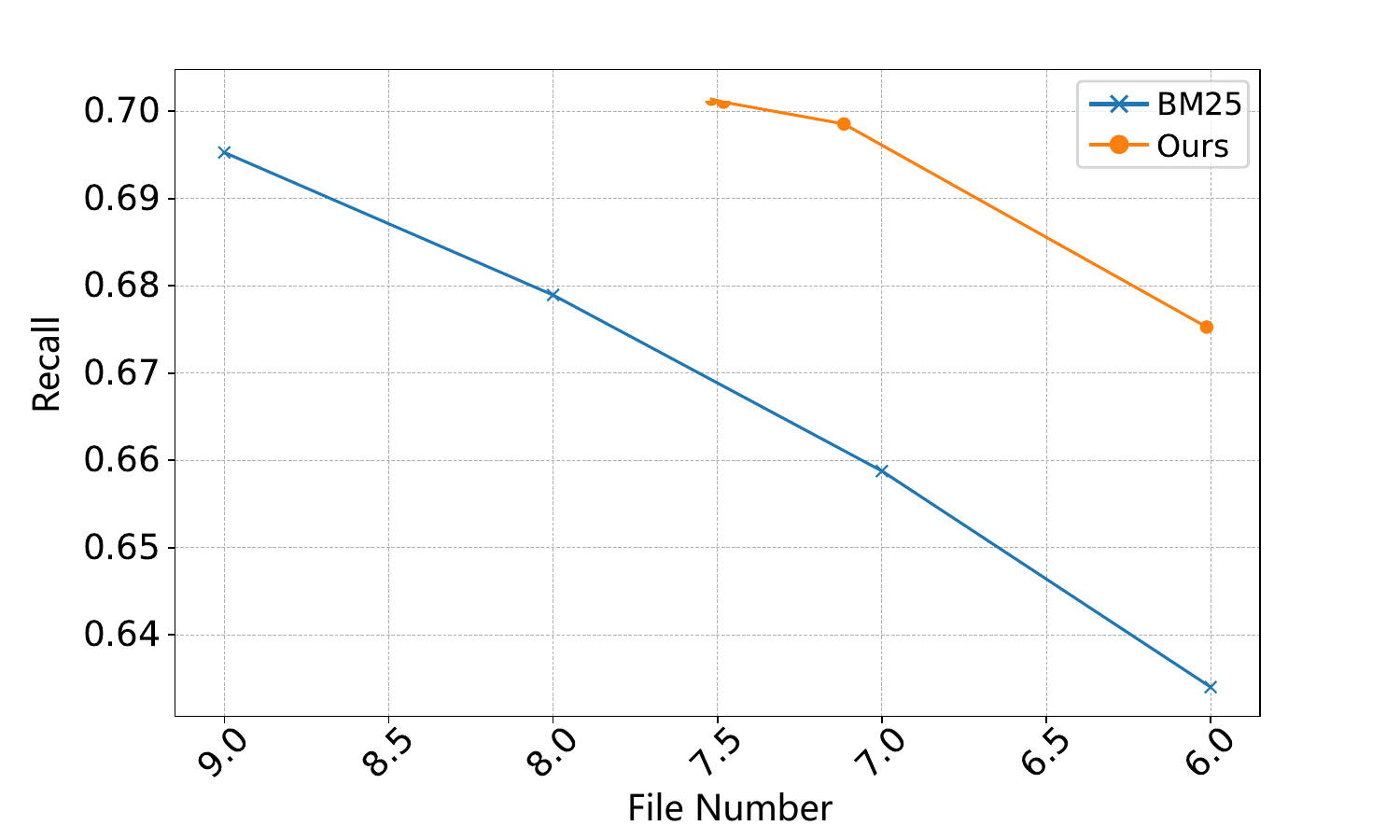}
        \vspace{-2em}
        \caption{\small Comparison of recall scores between Ours and BM25.}
        \label{fig:recall_FileNumber}
        \vspace{-0.8em}
    \end{minipage}
    \hfill
    \begin{minipage}[b]{0.52\textwidth}
        \centering
        \includegraphics[width=\linewidth]{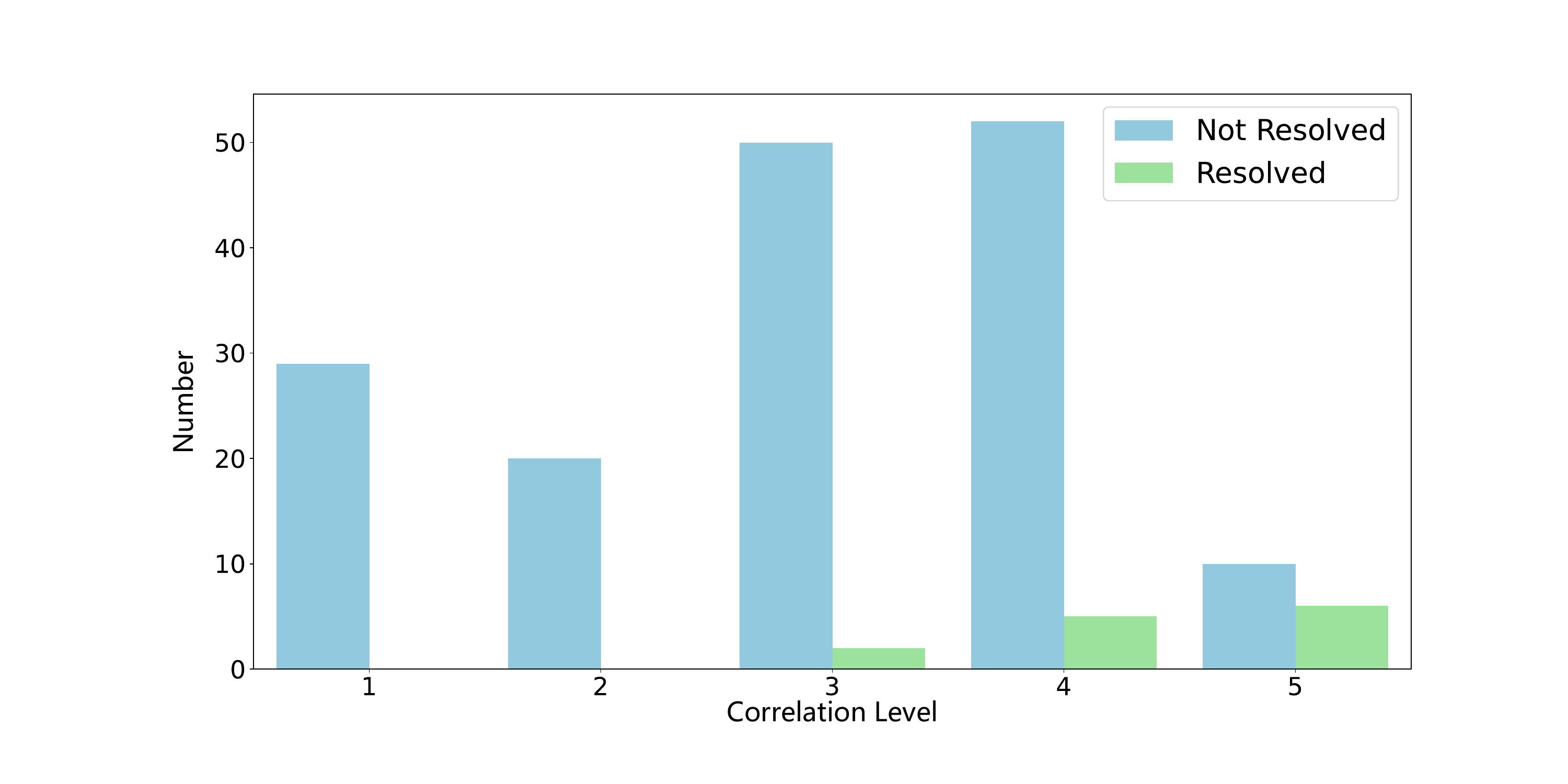}
        \vspace{-0.6cm}
        \caption{\small Distribution of the correlation score between the generated task description and the reference code change.}
        \label{fig:task_description_score_distribution}
        \vspace{-0.3cm}
    \end{minipage}
    \vspace{-1em}
\end{figure}

To investigate the effectiveness of the planning process, we analyze the Repository Custodian and Manager agent. 
The performance of the Repository Custodian agent is observed in the recall score versus the file number curve, as shown in \Fig~\ref{fig:recall_FileNumber}.
This curve demonstrates that our method consistently outperforms the BM25 baseline across varying numbers of selected files, indicating that our approach can identify the maximum number of relevant code files with the minimum selection.

For the Manager agent, we examined the alignment of its generated task descriptions with the reference code change by LLM. 
Following the study~\citep{DBLP:conf/nips/ZhengC00WZL0LXZ23}, we select \GptFour as an evaluator to score the correlation between the reference code change and the generated task description. 
The correlation scores are determined based on a set of criteria defined in \Tab~\ref{tab:score_meaning}.
A higher correlation score indicates a better alignment and thus, a more accurate and effective planning direction.
The distribution of these correlation scores is presented in \Fig~\ref{fig:task_description_score_distribution}.
Notably, most of the scores are $3$ or above, implying that the majority of task descriptions are in the right direction concerning planning. Furthermore, the higher scores correlate with a higher probability of issue resolution, indicated by a larger proportion of ``resolved'' outcomes in scores $4$ and $5$. This signifies that when the generated task description closely aligns with the reference, there is a higher possibility of resolving the issue.
The analysis above demonstrates the effectiveness of both the Repository Custodian and the Manager agent in the planning process of our framework.

\subsection{How Effective is Our Coding Process? (RQ 4)}\label{sec:rq4}

To evaluate the effectiveness of the coding process in our framework, we analyze the performance of Developers in locating code lines and resolving issues of different complexity.

\begin{figure}[t]
    \centering
    \begin{minipage}[b]{0.46\textwidth}
        \centering
        \includegraphics[width=\textwidth]{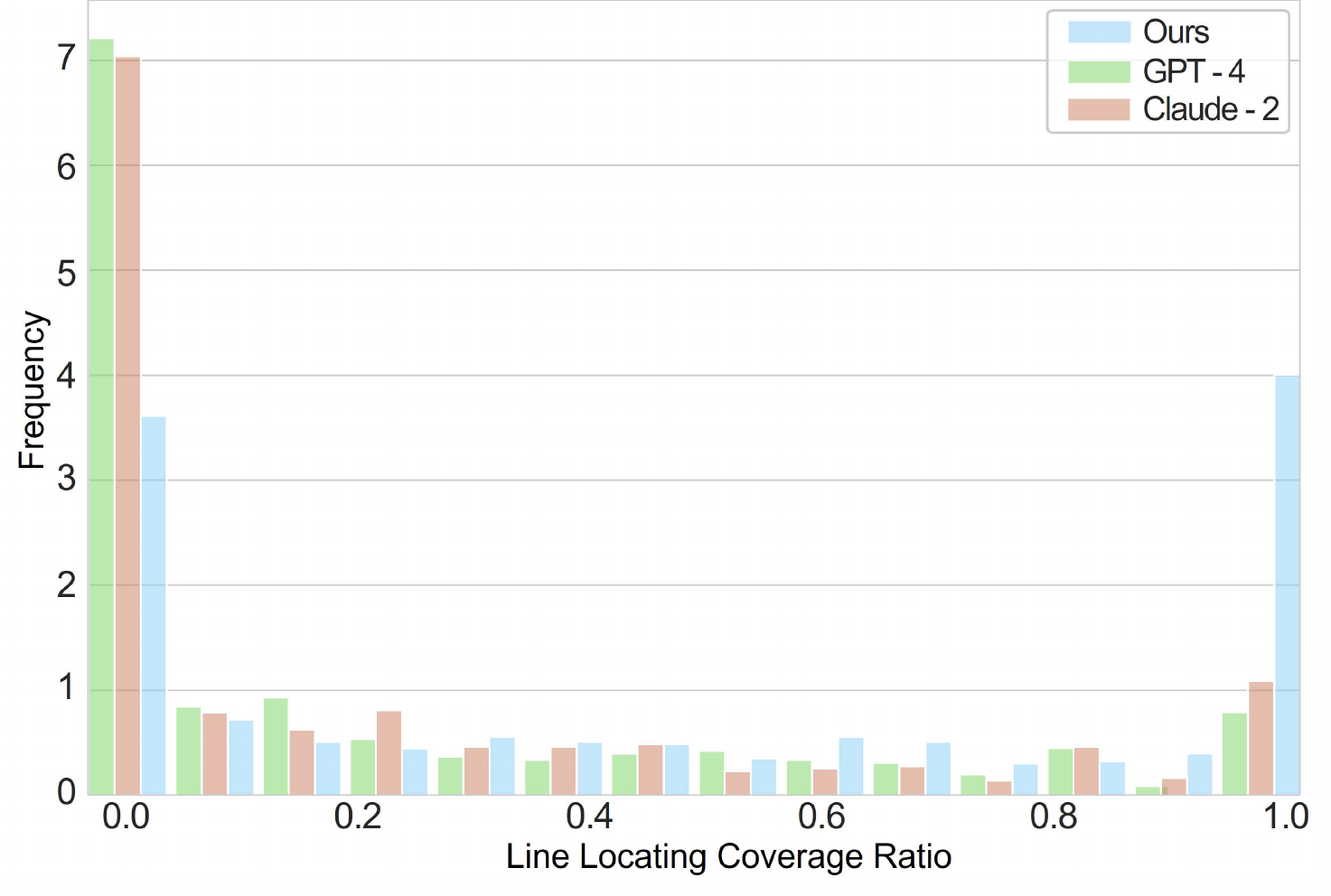}
        \vspace{-1em}
        \caption{\small Comparison of line locating coverage between \Ourmethod (Ours) and baselines.}
        \label{fig:line_location_overlap_ours}
    \end{minipage}
    \hfill
    \begin{minipage}[b]{0.53\textwidth}
        \centering
        \vspace{-2em}
        \includegraphics[width=\textwidth]{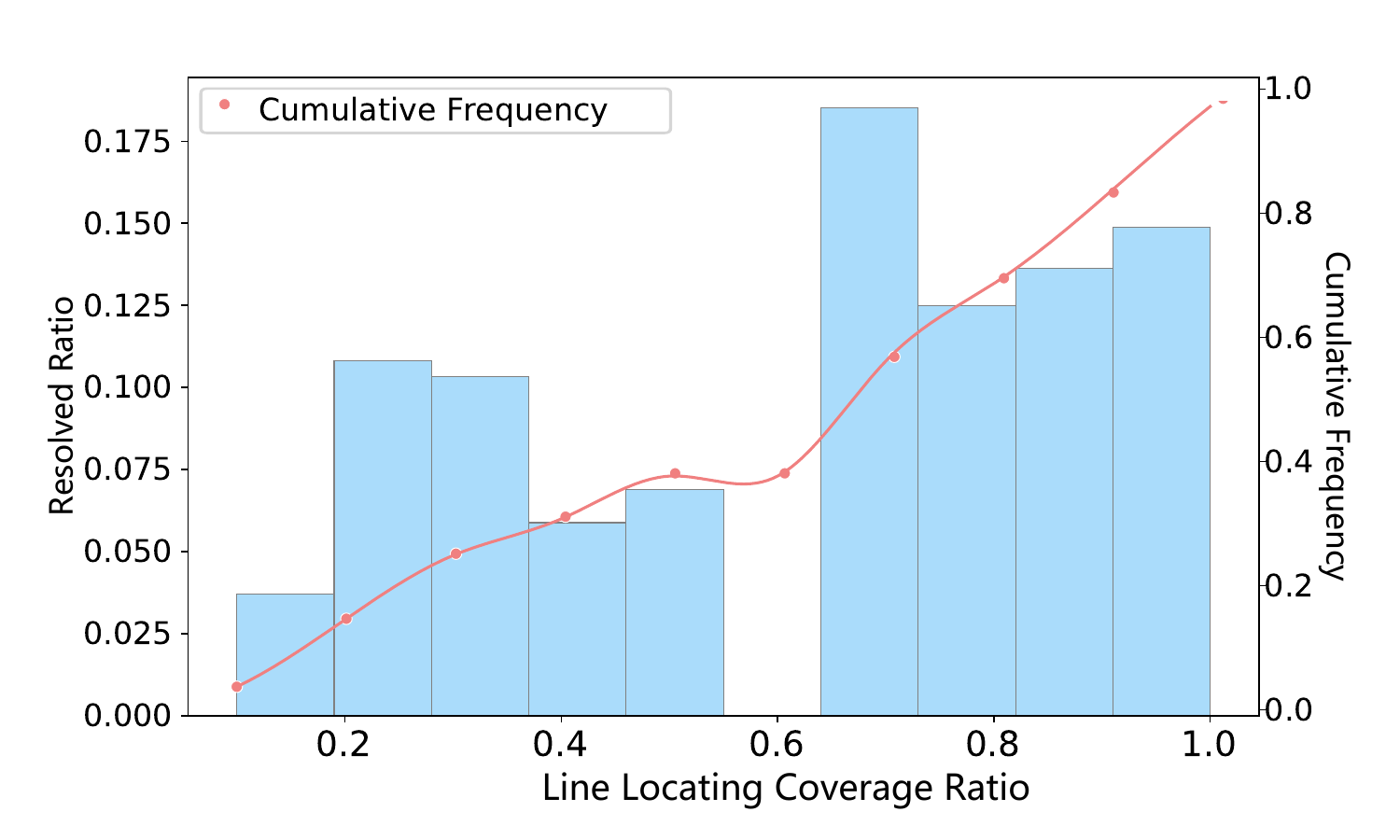}
        \vspace{-1.8em}
        \caption{\small Resolved ratio in different line locating coverage intervals.}
        \label{fig:interval_resolved_ratio}
    \end{minipage}
    \vspace{-2em}
\end{figure}

\Fig~\ref{fig:line_location_overlap_ours} illustrates the distribution of the line locating coverage ratio of \Ourmethod and the baselines.
This visualization reveals that our Developer agent frequently attains a line locating coverage ratio nearing $1$. Compared with baselines, the Developer agent demonstrates a pronounced preference for higher distribution values close to $1$, and conversely, a reduced preference for lower distribution values near $0$. Such a distribution validates the superior performance of \Ourmethod in locating code lines.

Further analysis is provided in \Fig~\ref{fig:interval_resolved_ratio} illustrating the relationship between the line locating coverage ratio and the issue resolved ratio within those coverages. 
As shown in \Fig~\ref{fig:interval_resolved_ratio}, the right four bars are higher than the five left, which indicates that the resolved ratio can increase with the line locating coverage. This observation also suggests that locating lines accurately is important for issue resolution. 
The cumulative frequency curve, shown in orange, provides an additional analysis, indicating the cumulative proportion of issues resolved ratio up to each point along the line locating coverage. 
A steady increase in cumulative frequency accompanies the increase in line locating coverage, reinforcing the idea that resolving issues is more successful in areas of high coverage. 
The slope of the curve's left half is lower than that of the right half, indicating that the benefits of increasing the coverage ratio are less pronounced at lower coverage ratios than at higher ones.
Therefore, the Developer agent should prioritize improving its capability of locating code lines.

Moreover, as shown in \Tab~\ref{tab:our_complexity_relation}, we present a logistic regression analysis that quantifies the correlation between several complexity indices and issue resolution. 
The results show that \GptFour has significant negative correlations across the number of files and functions, suggesting that as these indices increase, the likelihood of issue resolution decreases. Conversely, the negative correlations are less pronounced with our model, \Ourmethod, particularly in the number of files and functions, suggesting mitigation of challenges corresponding to these complexity indices. 

\begin{table}[ht]
    \footnotesize
    \centering
    \caption{\small Correlation between the complexity indices and the issue resolution.}
    \label{tab:our_complexity_relation}
    \begin{threeparttable}
    \begin{tabular}{lcccccc}
    \toprule  
    Method     & \# Files & \# Functions  & \# Hunks  & \# Added LoC    & \# Deleted LoC  & \# Changed LoC \\
    \midrule
    \GptFour	           & $-25.15$\tnote{*}	& $-25.15$\tnote{*}	&  $-0.06$\tnote{~}	&  $-0.10$\tnote{~}	&  $-0.04$\tnote{~}	&  $-0.21$\tnote{~}	\\
    \Ourmethod & \hspace{0.5em}$-1.55$\tnote{*}    & \hspace{0.5em}$-1.55$\tnote{*}    & $-0.12$\tnote{*}    & $-0.04$\tnote{*}    & $-0.06$\tnote{*}    & $-0.57$\tnote{*} \\
    \bottomrule
    \end{tabular}
    \begin{tablenotes}
    \item[*] The correlation between the index and the issue resolution is significant (P-value \(<\) $0.05$).
    \end{tablenotes}
    \end{threeparttable}
\end{table}

To evaluate the performance of the QA Engineer, the ablation experiment is conducted and the results are shown in \Tab~\ref{tab:swe_result_574}. As the table shows, in settings with and without hints, the presence of the QA Engineer can increase the resolved ratio by $1.57\%$ and $3.31\%$, respectively. This overall enhancement substantiates the QA Engineer's contribution to improving outcomes. Furthermore, a case detailed in Appendix~\ref{appendix:qa_case} underscores the QA Engineer's effectiveness.

\section{Related Work}

Researchers have developed LLM-based multi-agent systems, enabling more complex task completion. For instance, MetaGPT~\citep{hong2023metagpt,DBLP:journals/corr/abs-2402-18679} simulates a programming team's Standardized Operating Procedures (SOPs) and achieves leading scores on benchmarks like HumanEval~\citep{HumanEval} and MBPP~\citep{MBPP}. Similarly, ChatDev~\citep{qian2023communicative} functions as a virtual development company, decomposing requirements into atomic tasks and utilizing mutual communication and self-reflection to mitigate LLM hallucinations. While these systems excel in transforming requirements into code, they often overlook the challenges of code change generation during software evolution~\citep{DBLP:journals/corr/abs-2308-10620}.
\github issues include different types of requirements and most of them belong to bug fixing.
Previous researchers have proposed methods to localize the bugs~\citep{DBLP:conf/icse/ZhouZL12, DBLP:journals/tr/QiSYZM22} and some researchers explored various methods to automatic program repair\citep{DBLP:conf/icse/XiaWZ23,DBLP:journals/corr/abs-2403-17134,DBLP:conf/sigsoft/WongSKG21, DBLP:journals/toplas/AustinSF17, DBLP:conf/icse/YeM24, DBLP:conf/sigsoft/Wang0JH23}.
The full version of related work can be found in Appendix~\ref{appendix:related}. 

\section{Conclusion}

This paper illuminates the potential of LLMs in software development, particularly in resolving \github issues. Our empirical study identifies the challenges of direct LLM application. To address the challenges, we propose a novel LLM-based multi-agent framework, \Ourmethod, enhancing issue resolution through well-designed agents' collaboration. The superiority of \Ourmethod on the \Swebench against popular LLMs highlights its effectiveness, pointing towards a promising direction for integrating LLMs into software evolution workflows.

\bibliographystyle{plainnat}
\bibliography{reference}
\newpage

\appendix

\section{Detailed Explanation in Empirical Study}
\subsection{Coverage Ratio}\label{appendix:overlap_ratio}
The formula for calculating the coverage ratio is \Equ\ref{eq:overlap}.
As it shows, for each instance of \github issue resolution, the range of code change (in terms of the number of lines) in the reference $r$ is represented as a set of intervals $\mL_r = \{[s_{0},e_0], ..., [s_n,e_n]\}$, while the line ranges of the generated code change $g$ is $\mL_g = \{[s_0',e_0'], ..., [s_m',e_m']\}$, where $s$ and $e$ respectively represent the starting and ending line number of each modification hunk in the file, with $n$ hunks in the reference code change and $m$ hunks in the generated one.

\subsection{Observation on \Fig~\ref{fig:line_location_overlap_baselines}}\label{appendix:observation_line_location}
As shown in \Fig~\ref{fig:line_location_overlap_baselines}, we observe that: 
\ding{172} The distribution near the coverage ratio $0$ (left side of the figure) is the highest for all three LLMs, indicating that in most cases, the content generated by these models has a very low coverage ratio with the reference in terms of locating code lines. 
This means that these LLMs are most likely not able to accurately locate code lines that need to be modified in the process of generating the code change.
\ding{173} In the distribution near the line locating coverage of $1$ (right side of the figure), the three models show a consistent ranking (i.e., \ClaudeTwo $>$ \GptFour $>$ \ChatgptThreeFive) and this ranking is also consistent with the proportion of instances solved by the three models. 
This phenomenon suggests that the performance of LLMs in generating the code change is probably related to their ability to locate code lines accurately.

\subsection{Analysis on Complexity of the Code Change}\label{appendix:analysis_complexity}
As shown in \Fig~\ref{tab:empirical_complexity_relation}, compared with \ChatgptThreeFive and \GptFour, \ClaudeTwo exhibits a different pattern, with much lower negative correlations for the number of files and functions, which indicates that it is a more efficient approach to generate the code change for \github issue resolution. However, it also shows significant negative correlations across other indices such as the number of hunks, added lines of code (LoC), deleted LoC, and changed LoC.

\section{Kick-off Meeting Example}\label{appendix:meeting}

\begin{figure}
    \centering
    \includegraphics[width=\linewidth]{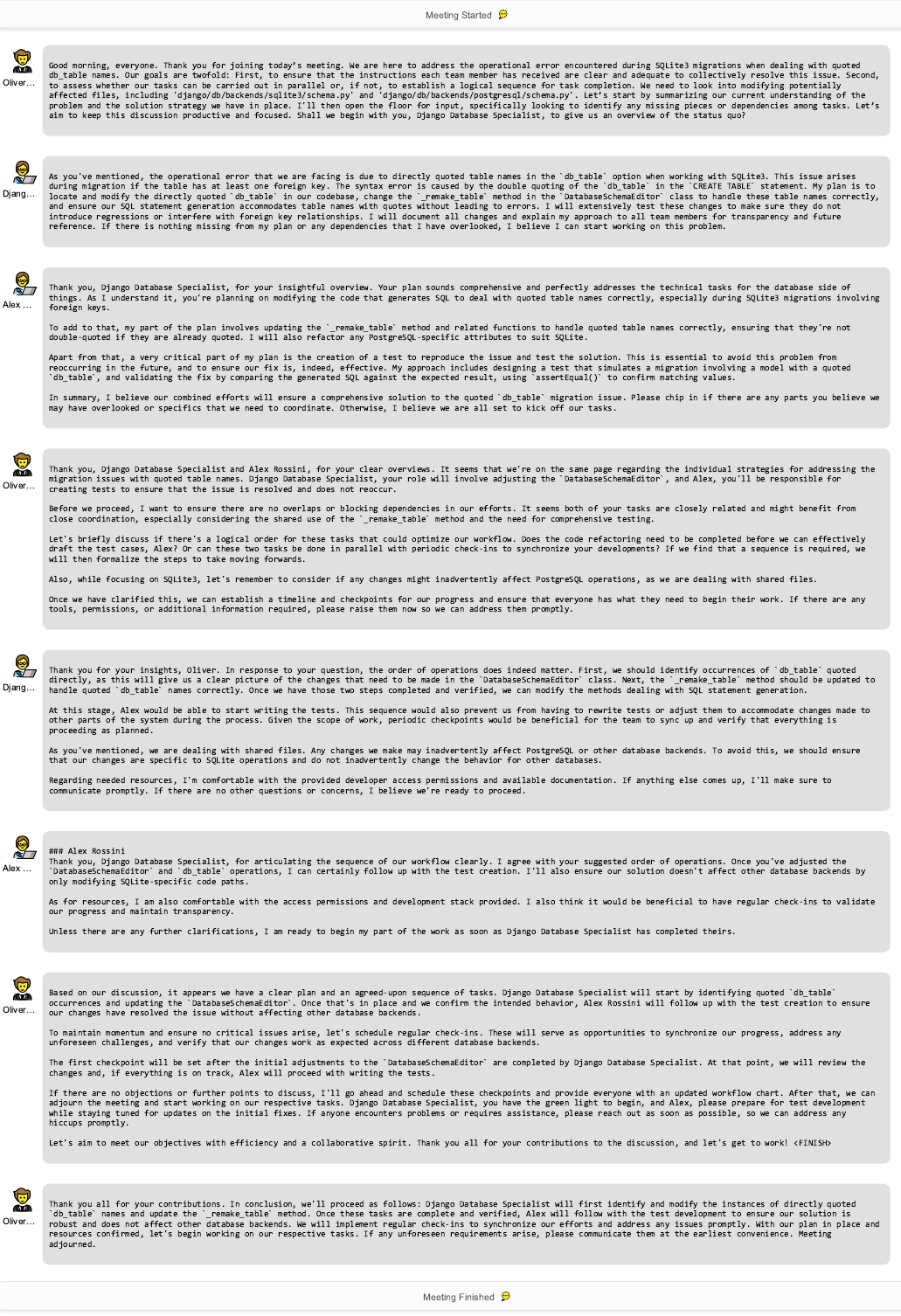}
    \caption{Kick-off meeting to resolve the issue~\citep{django30664}.}
    \label{fig:kick_off_meeting}
\end{figure}

Figure~\ref{fig:kick_off_meeting} illustrates a kick-off team meeting. In this meeting, three participants are present: the Manager agent, Oliver CodeLead, and two Developer agents, Django Database Specialist and Alex Rossini. They discuss a specific issue\footnote{\url{https://code.djangoproject.com/ticket/30664}}, assigned tasks, and determine the workflow sequence.

\section{Metrics}\label{appendix:metric}

The applied ratio indicates the proportion of instances where the code change is successfully generated and can be applied to the existing code repository using Git tools, i.e.,
\begin{align}
    \text{Applied Ratio} = \frac{\left| \gD  \right|}{\left| \gI \right|}, ~\label{eq:applied_ratio}
\end{align}
where $\gD$ represents the set of instances in the generated code change set that could be applied to the original code repository using the ``\texttt{git apply}'' operation, and $\gI$ is the set of all instances in the test set.
The resolved ratio refers to the proportion of instances in which the code change is successfully applied and passed a series of tests, i.e.,
\begin{align}
    \text{Resolved Ratio} = \frac{\left| \sum_{i=0}^{l} (\{T_{old}( d_i)\} \cap \{T_{new}( d_i)\} )\right|}{\left| \mathcal{I} \right|}, ~\label{eq:resolved_ratio}
\end{align}
where $T_{old}$ denotes all the test cases that the old version of the code repository could pass, $T_{new}$ represents all the test cases designed for new requirements, and $d_i$ denotes the code change generated to resolve the issue in the $i$-th instance. Furthermore, $T(d) = \True$ means that the code change $d$ can pass all the test cases in $T$.

The recall score versus file number curve is used to measure the effectiveness of locating code files to be modified. The recall score refers to the proportion of files that are successfully located out of all the files that require modification. The formula for calculating the file locating recall score for the $i$-th instance is as follows:
\begin{align}
\text{Recall} = \frac{\left| \gG_i \cap \gR_i \right|}{\left| \gR_i \right|} \times 100\%, ~\label{eq:recall_file}
\end{align}
where $\gG_i = \sum_{j=0}^{n}{g_{i,j}}$ represents the set of file paths located by our framework, with each file path in the set denoted as $g_{i,j}$ and the total number of files as $n$;
$\gR_i = \sum_{k=0}^{m}{r_{i,k}}$ denotes the paths of the files that need to be modified, with each reference file path denoted as $r_{i,k}$ and the total file number as $m$.
In this curve, ``file number'' refers to the average number of files that need to be processed across all instances to achieve the given recall score. 
Specifically, it illustrates how many files averagely need to be located by our framework before reaching the recall score denoted by the curve at any point. This metric represents both the effectiveness and efficiency of file locating.

\section{Comparison Result on \Swebench Lite}\label{appendix:exp_lite}

Recently, some contemporaneous works~\citep{opendevin2024}, e.g., AutoCodeRover~\citep{DBLP:journals/corr/abs-2404-05427} and SWE-Agent~\citep{yang2024sweagent}, have been proposed for this task. These methods are evaluated using \Swebench lite, a canonical subset of \Swebench, which is recommended for evaluation~\citep{swebenchlite}. Considering budget constraints, we conducted experiments on \Swebench lite to compare with them on the same issues' resolution.

The experimental results are shown in \Tab~\ref{tab:swe_result_lite}. \Ourmethod achieves the highest resolved ratio, $25.33\%$, than other baselines. The performance of \Ourmethod slightly decreased when evaluated without QA, reaching $23.33\%$, and dropped under the other two ablation settings. This comparative study underscores the robustness of \Ourmethod, particularly when provided with comprehensive inputs, and highlights the impact of QA and hints on its performance. The results indicate that while new methods like AutoCodeRover and SWE-Agent show promise, \Ourmethod remains an effective method for \github issue resolution.

\begin{table}[ht]
\small
    \centering
    \caption{\small The comparison of overall performance between \Ourmethod and baselines on \Swebench lite.}
    \label{tab:swe_result_lite}
    \begin{threeparttable}
    \begin{tabular}{l|cccccc}
    \toprule
    \multirow{2}{*}{\textbf{Method}}         & \multirow{2}{*}{AutoCodeRover} & \multirow{2}{*}{SWE-Agent} & \multicolumn{4}{c}{\Ourmethod} \\
    \cmidrule(r){4-7}
    &&& Full & w/o QA   & w/o hints & w/o (hints, QA) \\
    \midrule
    \textbf{Resolved}    & $16.11\%$ ($22.33\%$*) & $18.00\%$     & $\textbf{25.33\%}$
    & $23.33\%$                 & $16.67\%$         & $16.00\%$     \\
    \bottomrule
    \end{tabular}
    \begin{tablenotes}
    \item[*] {\footnotesize Note that $16.11$ is the average scores among 3 runs while $22.33$ is under the union of from the 3 runs.}
    \end{tablenotes}
    \end{threeparttable}
\end{table}

\section{Comparison with Devin}\label{appendix:devin}

Devin is a novel agent for software development~\citep{devin}, and its performance has also been assessed using the \Swebench. However, the evaluation dataset employed by Devin differs from the subset used for experiments with \GptFour reported by the paper of \Swebench~\citep{jimenez2024swebench}. An analysis of the repository name and pull request ID of each instance reveals that only $140$ instances coverage between the two datasets.

Within the shared pool of $140$ instances, our framework successfully resolves $21$ ($15\%$) issues, surpassing Devin's resolution of $18$ ($12.86\%$) issues~\footnote{\url{https://github.com/CognitionAI/devin-swebench-results/tree/main/output_diffs/pass}}. This comparison, however, may not be entirely equitable. Devin's possible underlying LLM is unknown, and it possesses the capability to integrate feedback from the environment. Moreover, Devin's reported scores are under the setting given the entire repository, and it operates with ``common developer tools including the shell, code editor, and browser'', and ``agents with internet access could potentially find external information through other methods'' as detailed at the report~\footnote{\url{https://www.cognition-labs.com/introducing-devin}}. In contrast, our approach solely relies on the shell, without the need of any additional external tools.

For running time, 72\% of instances resolved by Devin require greater than $10$ minutes to complete.
In contrast, our framework finalizes each resolved issue within approximately 3 minutes.
On average, our framework completes the processing of each instance in under 5 minutes, demonstrating its capability to assist in resolving \github issues with minimal time expenditure.

\section{Statistics on the Generated Code Changes}\label{appendix:stat}

This section provides statistics on code changes corresponding to resolved issues and those applicable but unresolved using our framework.

The statistics on the code change for instances with resolved issues are presented in \Tab~\ref{tab:stats_combined}. 
Overall, the statistical information of the generated code changes for these instances, such as the average number of code files, functions, hunks, and deleted lines, all differ slightly (not exceeding $0.3$) from the reference solutions written by humans. This indicates that for these instances, the complexity of the code change generated by our framework is similar to that of humans. Furthermore, the maximum values observed in the table reveal that our framework can implement code modifications involving two files, four hunks, and $1,655$ lines modification, with single modifications reaching up to $190$ lines. Results demonstrate the effectiveness of our method in resolving complex issues that need to modify the code file on multiple locations and with long context.

\begin{figure}[b]
    \centering
    \begin{minipage}[b]{0.52\textwidth}
        \centering
        \includegraphics[width=0.48\textwidth]{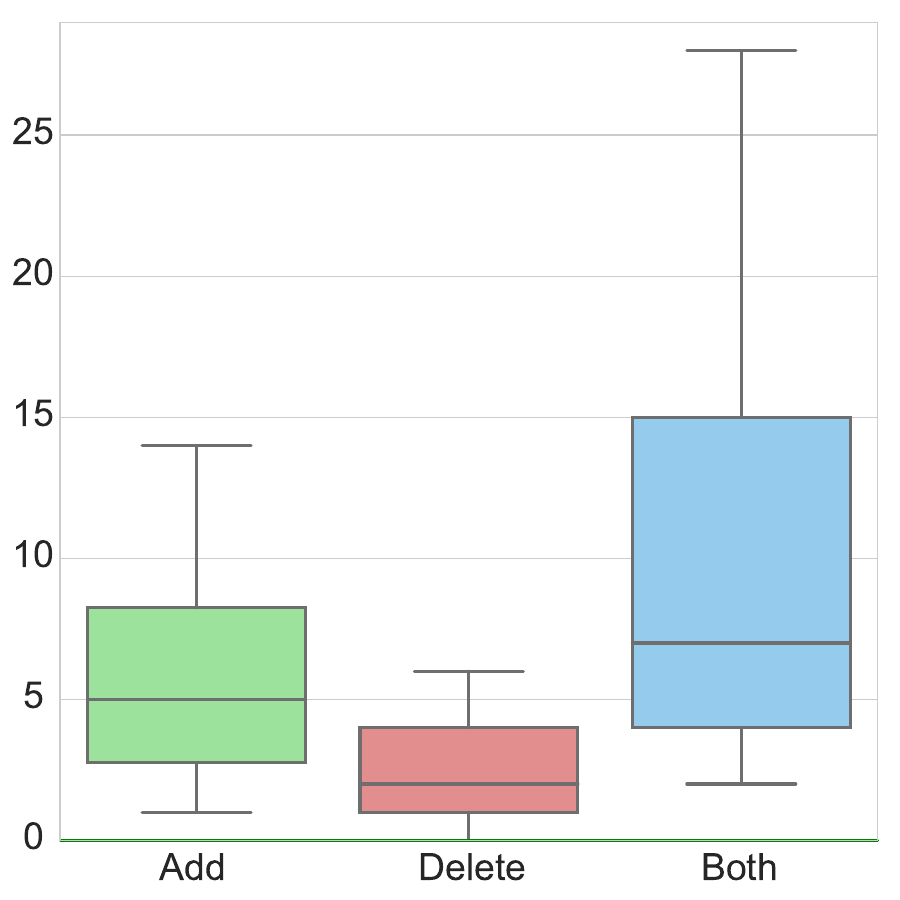}
        \includegraphics[width=0.48\textwidth]{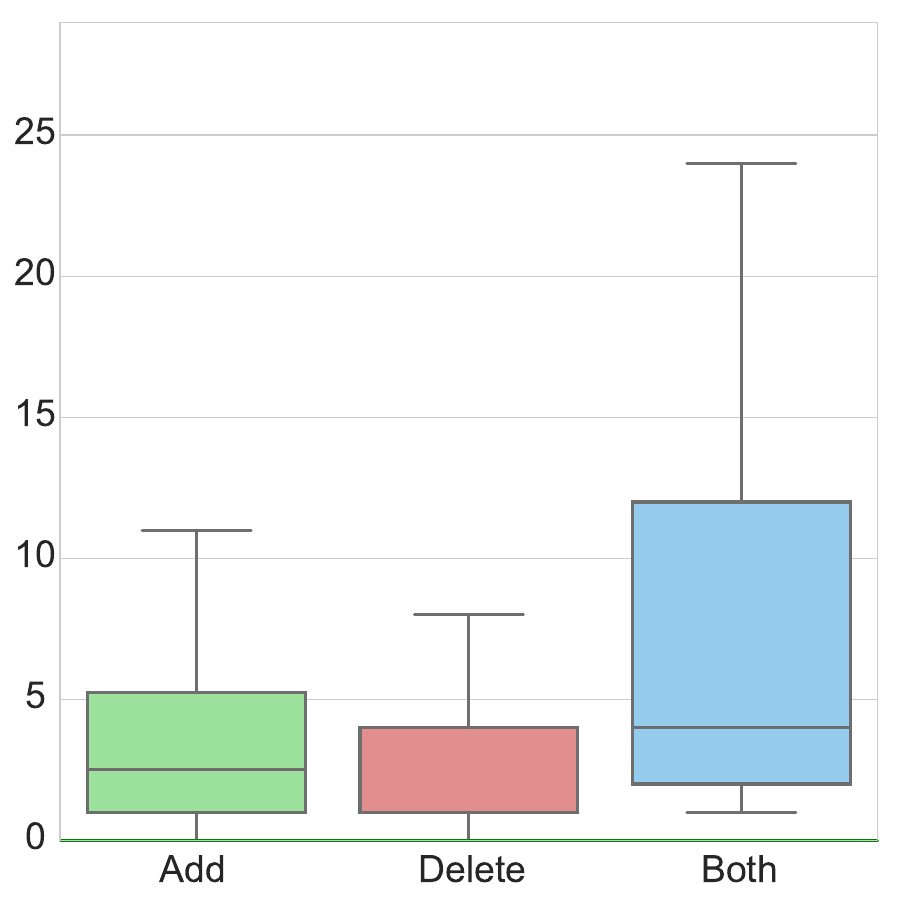}
        \caption{\small Distribution of the LoC in the resolved instances.}
        \label{fig:resolved_part_change_col_dis}
    \end{minipage}
    \hfill
    \begin{minipage}[b]{0.46\textwidth}
        \centering
        \includegraphics[width=0.48\textwidth]{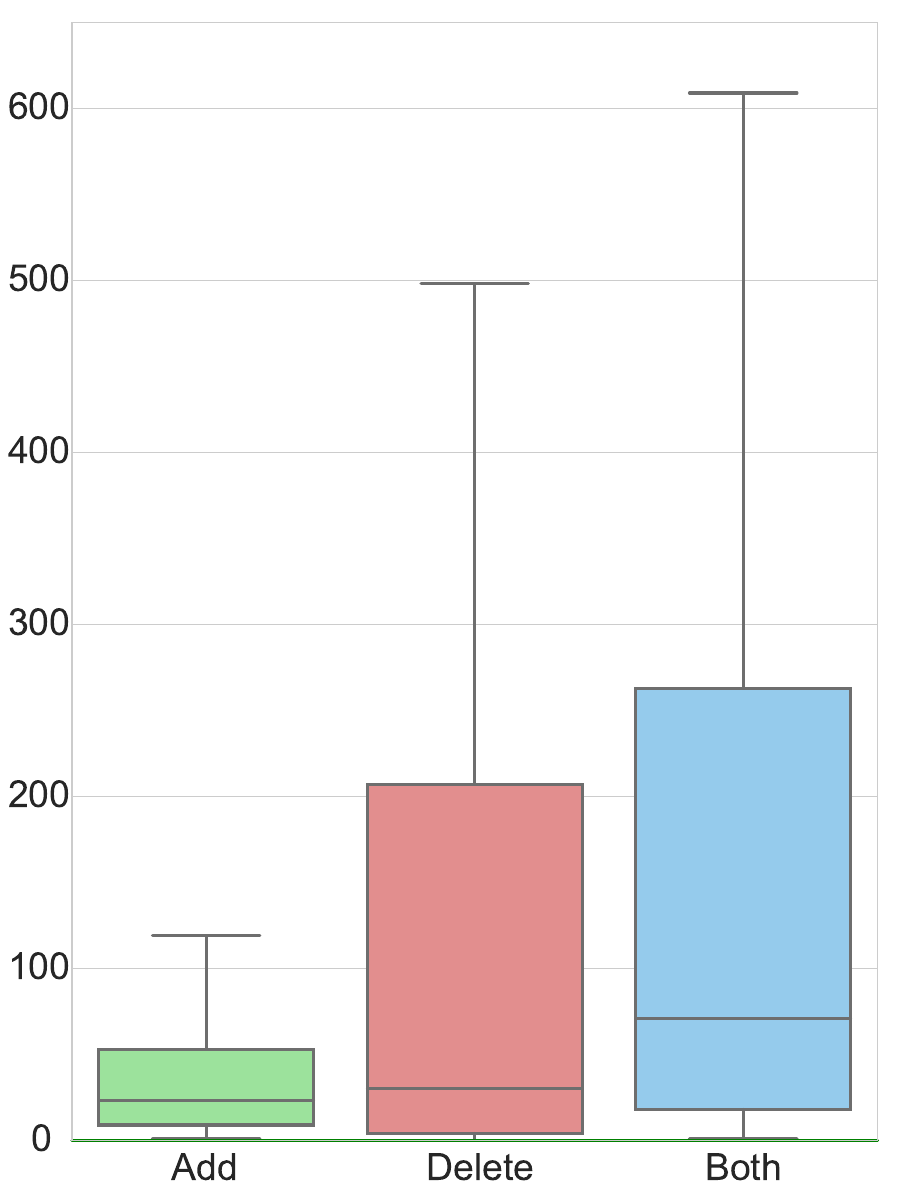}
        \includegraphics[width=0.48\textwidth]{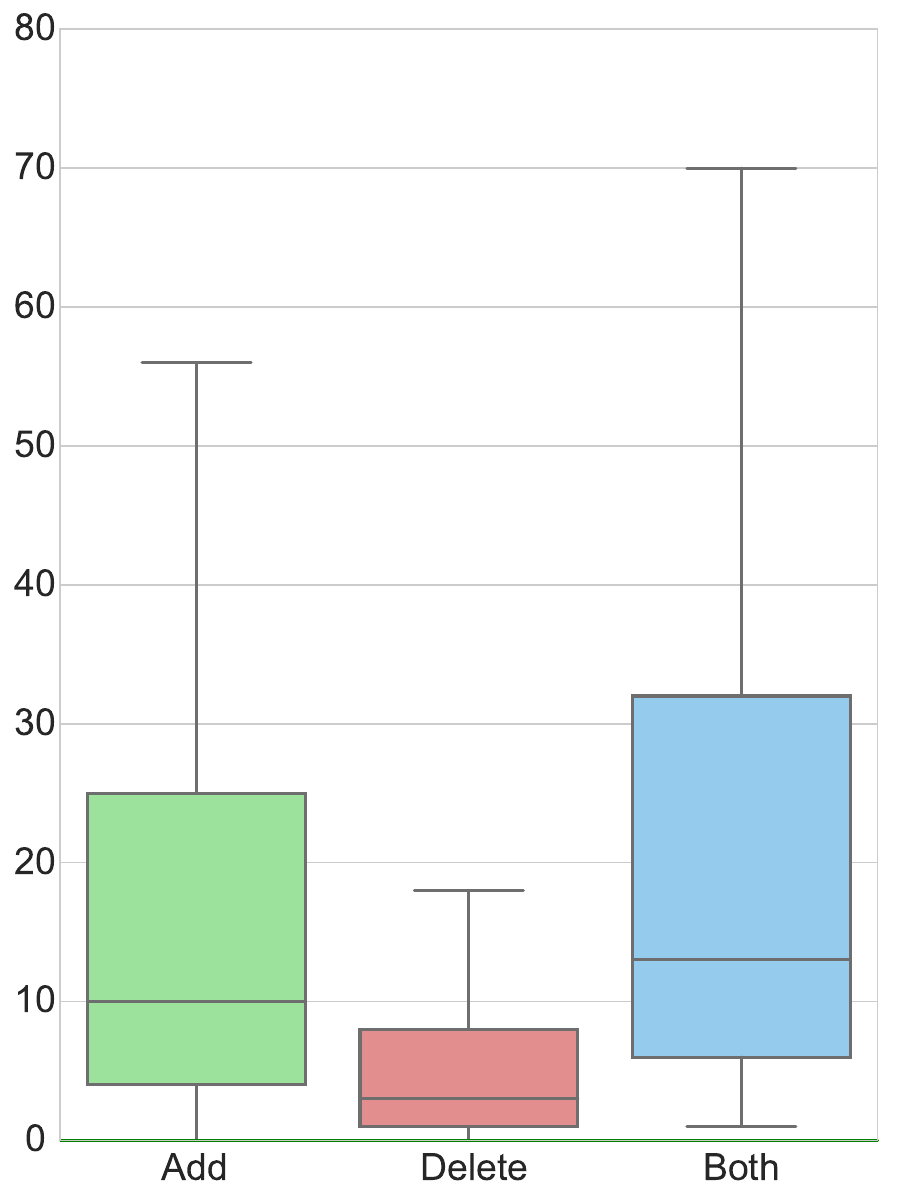}
        \caption{\small Distribution of the LoC in the applied but not resolved instances.}
        \label{fig:applied_part_change_col_dis}
    \end{minipage}
    \vspace{3em}
\end{figure}

\begin{figure}[ht]
    \centering
    \begin{minipage}[b]{0.49\linewidth}
        \centering
        \includegraphics[width=\linewidth]{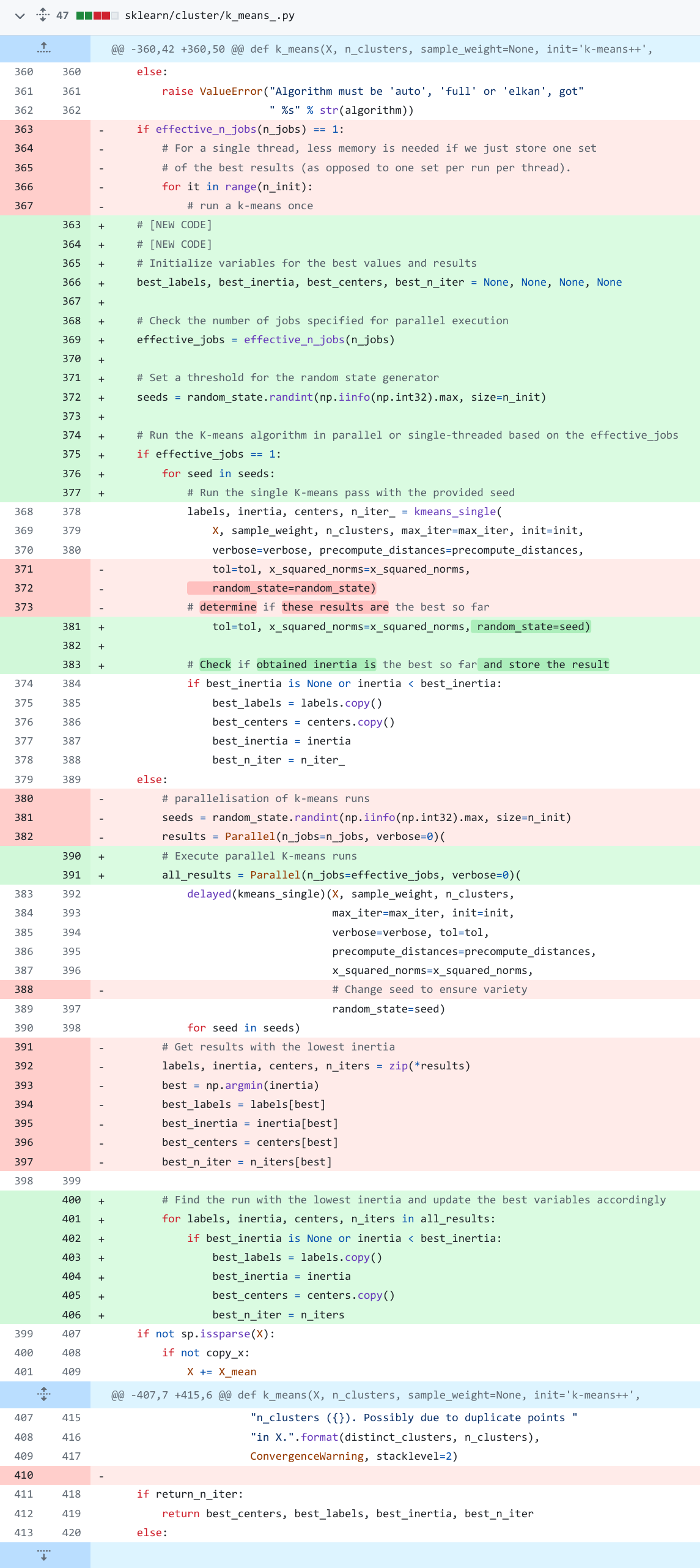}
        \caption{\small Case from \texttt{scikit-learn} (ours, after review) for the issue~\citep{sklearn9784}.}
        \label{fig:sklearn_case_our_after_review}
    \end{minipage}
    \hfill
    \begin{minipage}[b]{0.48\linewidth}
        \centering
        \includegraphics[width=\linewidth]{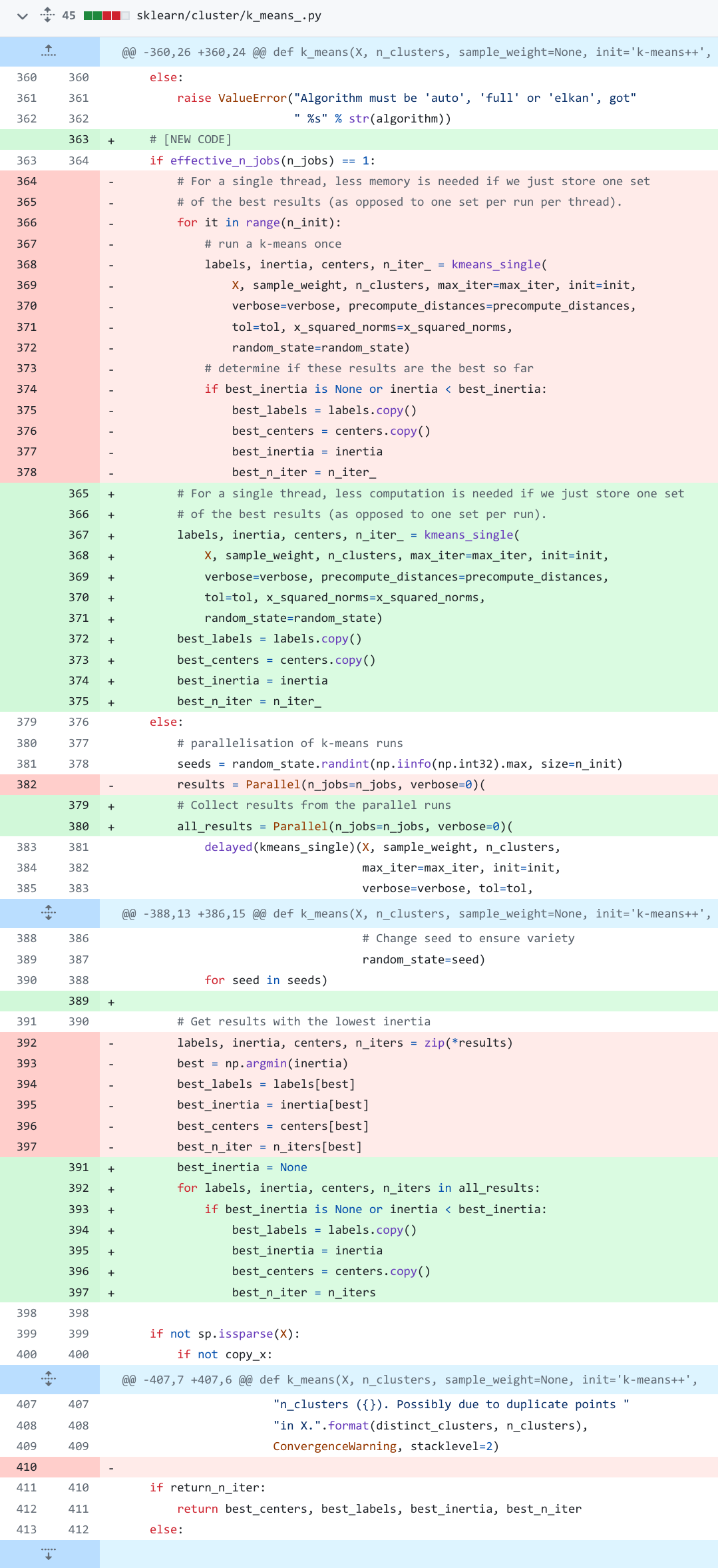}
        \caption{\small Case from \texttt{scikit-learn} (ours, before review) for the issue~\citep{sklearn9784}.}
        \label{fig:sklearn_case_our_before_review}
    \end{minipage}
\end{figure}

\begin{figure}[ht]
    \centering
    \begin{minipage}[b]{0.49\linewidth}
        \centering
        \includegraphics[width=\linewidth]{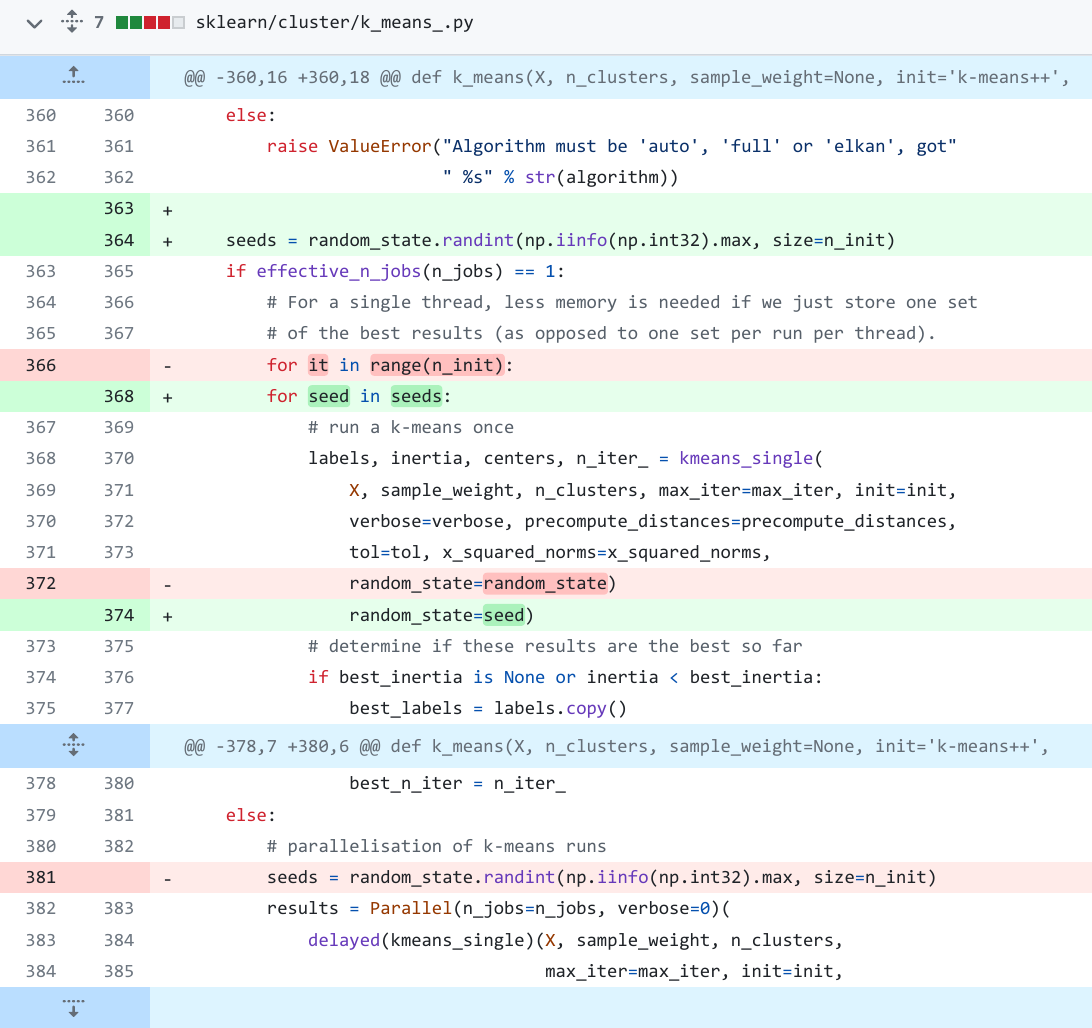}
        \caption{\small Case from \texttt{scikit-learn} (gold)~\citep{sklearn9288}.}
        \label{fig:sklearn_case_gold}
    \end{minipage}
    \hfill
    \begin{minipage}[b]{0.48\linewidth}
        \centering
        \includegraphics[width=\linewidth]{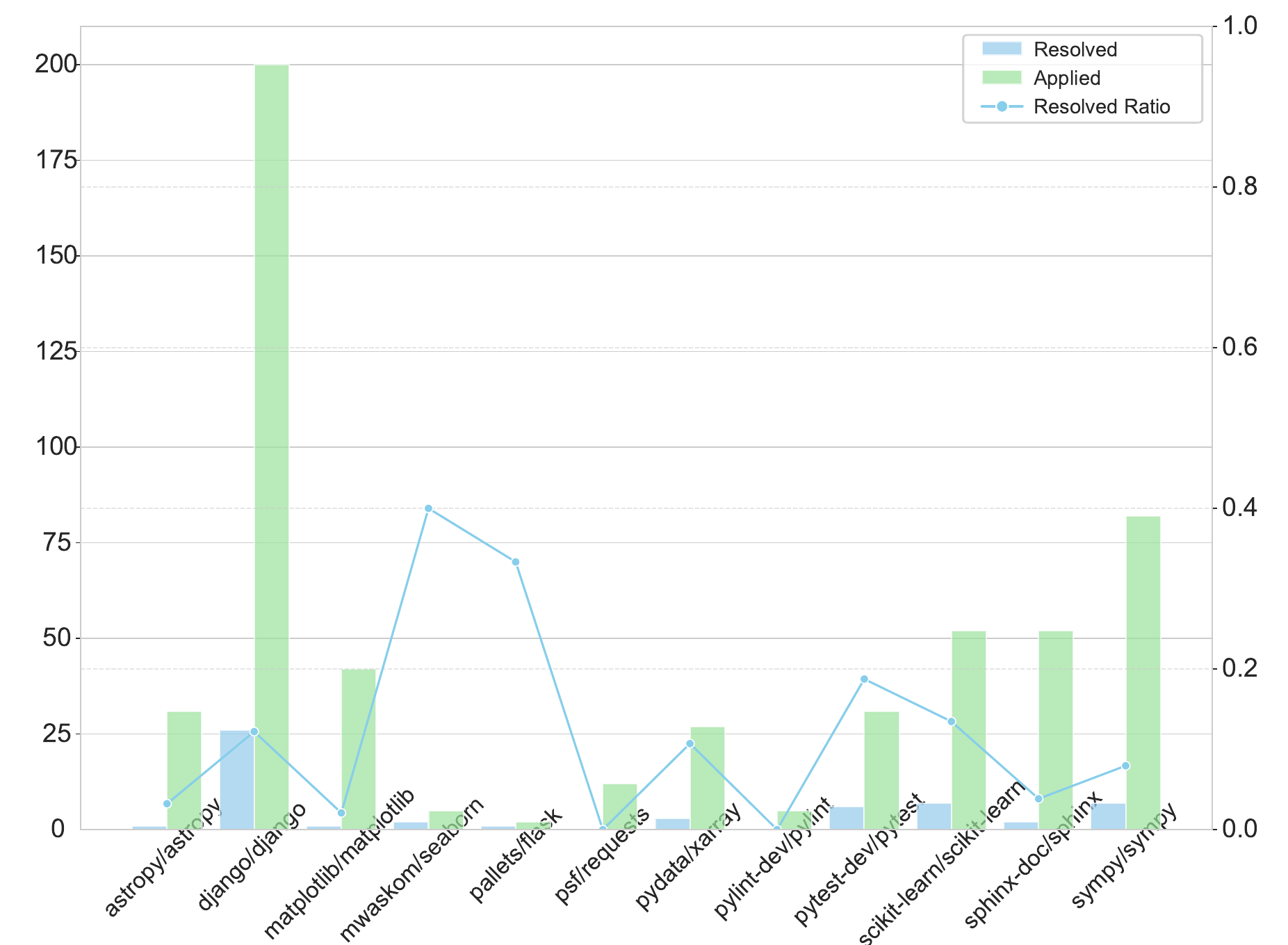}
        \caption{\small The number of applied and resolved instances in different repositories.}
        \label{fig:applied_resolved_ratio_per_repo}
    \end{minipage}
\end{figure}

Specifically, the distribution of the number of modified lines for the resolving instances is shown in \Fig~\ref{fig:resolved_part_change_col_dis}. We observe that the distribution of the number of modified lines in our framework for the solved instances exceeds that of the reference solution, especially in terms of the number of added lines being significantly higher than the reference. 
Upon manual inspection, we found that the generation results provided by our framework often contained more comment information, which led to an increase in the total number of modified lines. For example, \Fig~\ref{fig:sklearn_case_our_after_review} displays the generation result of our framework. Lines $365, 368, 371, 374, 383$ in the new version file correspond to the comment for the added code. These natural language descriptions are valuable in actual software evolution~\citep{DBLP:conf/icse/HuX0WCZ22, DBLP:conf/icse/MuCSWW23}. In contrast, \Fig~\ref{fig:sklearn_case_gold} shows a human-written solution lacking such explanatory comments, which might disadvantage software maintainers in reading and understanding.

The statistics on the code change for instances without resolved issues are shown in \Tab~\ref{tab:stats_combined}. From the table, our framework can generate applicable code changes including up to $13$ files and $28$ hunks, and the location of the modifications can be as far as line $7,150$, with a single modification reaching up to $9,367$ lines. These results suggest that our method has a strong adaptability in generating applicable code changes. However, considering that these code changes have not passed all the potential test cases they could pass, which indicates that there is still room for improvement.

To further analyze the reasons behind the failure of test cases in these instances, we have quantified the distribution of the lengths of code changes in the unresolved instances, as shown in \Fig~\ref{fig:applied_part_change_col_dis}. From the figure, we observe that for unresolved instances, the framework tends to delete a larger number of lines while adding fewer lines, in contrast to the distribution of human-written changes. 
This discrepancy may point to different repair strategies or attitudes towards problem-solving, where the framework presented herein might prefer to reduce errors by removing potentially problematic code, whereas human developers may lean towards adding new code to address issues.

Moreover, a comparison between the resolved instances and not resolved ones shown in \Tab~\ref{tab:stats_combined} reveals that the latter contains a higher overall number of files, hunks, and changed lines of code. These instances, involving more modification locations, correspond to more complex scenarios. This phenomenon suggests that the performance of our framework in resolving such complex issues requires further enhancement.

Furthermore, the variability in difficulty across different software repositories may influence the effectiveness of code changes. To this end, we compile statistics on the resolved ratios in various software repositories, as shown in \Fig~\ref{fig:applied_resolved_ratio_per_repo}. From the figure, we observe that there is a significant variation in the resolved ratios across different repositories in our framework. Some repositories have a resolved ratio as high as $40\%$, while others are close to $0\%$. This suggests that the differences among various software such as code structure and coding style can impact the generation and application of the code change.

\begin{table}[ht]\scriptsize
    \centering
    \caption{\small The statistical analysis of our framework on resolved and applied but not resolved instances.}
    \label{tab:stats_combined}
    \begin{tabular}{lrrr|rrr|rrr|rrr}
    \toprule  
    & \multicolumn{6}{c}{\textbf{Resolved Instances}} & \multicolumn{6}{c}{\textbf{Applied but Not Resolved Instances}} \\
    \cmidrule(r){2-7} \cmidrule(r){8-13}
    & \multicolumn{3}{c}{\textbf{\Ourmethod}} & \multicolumn{3}{c}{\textbf{Gold}} & \multicolumn{3}{c}{\textbf{\Ourmethod}} & \multicolumn{3}{c}{\textbf{Gold}} \\
    \cmidrule(r){2-4} \cmidrule(r){5-7} \cmidrule(r){8-10} \cmidrule(r){11-13}
    & {\textbf{Min}} & {\textbf{Max}} & {\textbf{Avg.}} & {\textbf{Min}} & {\textbf{Max}} & {\textbf{Avg.}} & {\textbf{Min}} & {\textbf{Max}} & {\textbf{Avg.}} & {\textbf{Min}} & {\textbf{Max}} & {\textbf{Avg.}} \\
    \midrule
    \# Code Files            & 1 & 2 & 1.02 & 1 & 2 & 1.04 & 1 & 13 & 1.50 & 1 & 18 & 1.61 \\  
    \# Functions             & 1 & 2 & 1.02 & 1 & 2 & 1.04 & 1 & 13 & 1.50 & 1 & 18 & 1.61 \\
    \# Hunks                 & 1 & 4 & 1.45 & 1 & 6 & 1.66 & 1 & 28 & 2.52 & 1 & 52 & 3.72 \\
    \# Added Lines           & 1 & 146 & 9.75 & 0 & 38 & 4.34 & 1 & 920 & 40.38 & 0 & 3,050 & 28.27 \\
    \# Deleted Lines         & 0 & 77 & 5.27 & 0 & 115 & 5.16 & 0 & 9,160 & 327.27 & 0 & 2,975 & 14.51 \\
    Change Start Index       & 1 & 1,655 & 270.12 & 1 & 1,657 & 256.09 & 1 & 4,568 & 424.84 & 0 & 6,651 & 485.01 \\
    Change End Index         & 22 & 1,665 & 301.68 & 0 & 1,666 & 315.05 & 9 & 7,150 & 513.13 & 0 & 6,658 & 728.96 \\ 
    \# Changed Lines         & 2 & 190 & 15.02 & 1 & 115 & 9.50 & 1 & 9,367 & 367.65 & 1 & 6,025 & 42.79 \\ 
    \bottomrule
    \end{tabular}
\end{table}

\section{Evaluation on Task Description}\label{appendix:eval_task_desc}

Since there is no ground truth for the task descriptions generated by the Manager, we utilize \GptFour to simulate human evaluation and score each task description based on its corresponding reference code change. Table \ref{tab:score_meaning} illustrates the standards used by \GptFour to assess the correlation between the task description and the code change. The score given by \GptFour is considered the performance metric for the task description.

\begin{table}[ht]\footnotesize
\centering
\caption{\small The meaning of scores in \GptFour evaluation on the correlation between the generated task description and the reference code change.}
\begin{tabular}{cl} \toprule
\textbf{Score}   & \multicolumn{1}{c}{\textbf{Meaning}}  \\ \midrule 
1       & The code changes are unrelated to the task description. \\\addlinespace[0.2em]
2       & The code changes address a minor part of the task but are largely irrelevant. \\\addlinespace[0.2em]
3       & The code changes partially meet the task requirements but lack completeness or accuracy. \\ \addlinespace[0.2em]
\multirow{2}{*}{4}  & The code changes are relevant and mostly complete, with minor discrepancies from the\\ 
& task description. \\\addlinespace[0.2em]
\multirow{2}{*}{5}       & The code changes perfectly align with the task description, fully addressing all specified \\
& requirements with high accuracy and completeness. \\ \bottomrule
\end{tabular}
\vspace{-1em}
\label{tab:score_meaning}
\end{table}

\section{Case Study}\label{appendix:case_study}

\begin{figure}
    \centering
    \includegraphics[width=\linewidth]{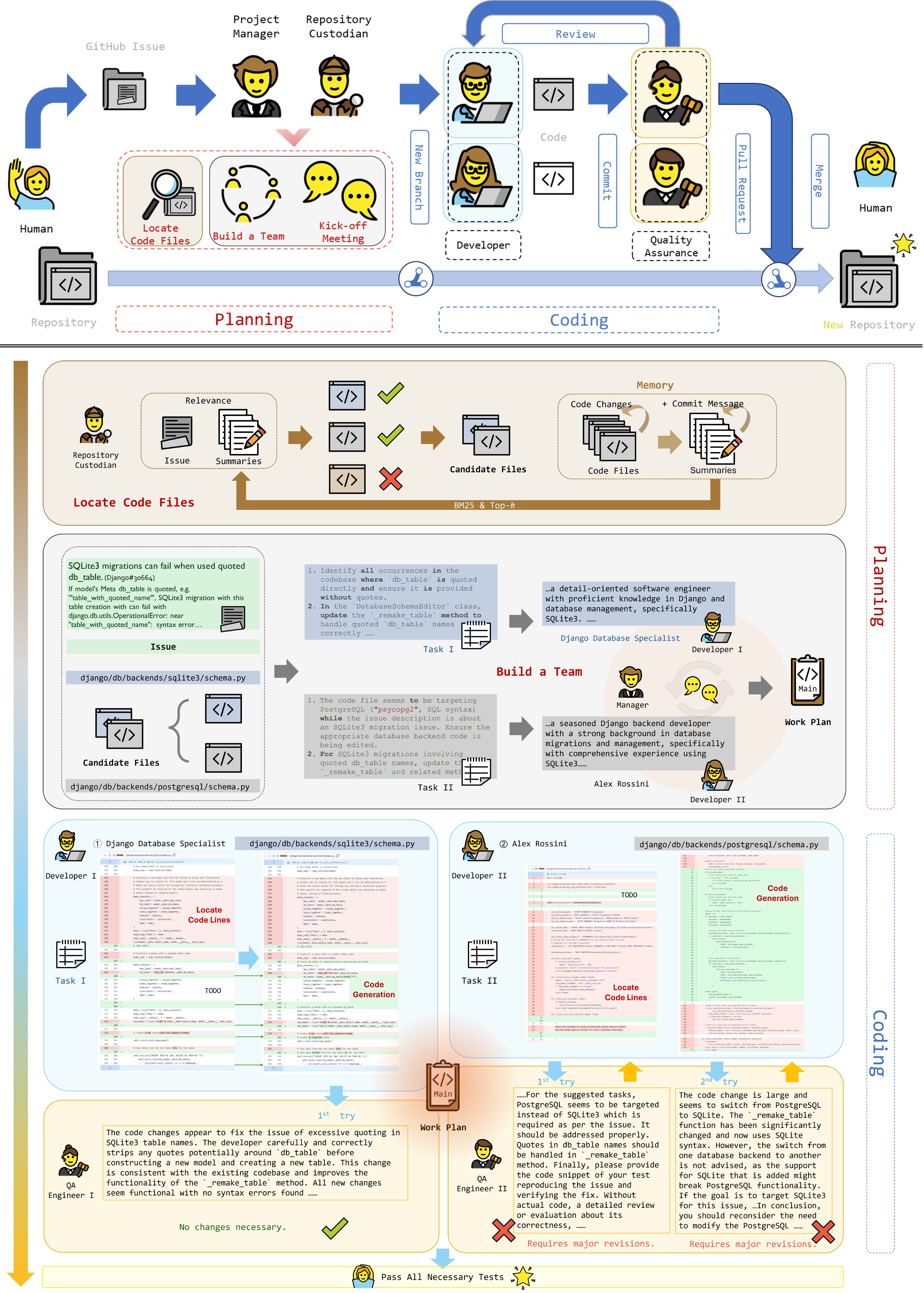}
    \caption{\small Detailed overview of our framework, \Ourmethod (Kick-off meeting refers to \Fig\ref{fig:kick_off_meeting}).}
    \label{fig:detailed_overview}
\end{figure}

\Fig\ref{fig:detailed_overview} illustrates the detailed process of our framework used to resolve the issue from the \texttt{Django} repository~\citep{django} as described in the following ticket~\footnote{\url{https://code.djangoproject.com/ticket/30664}}.
To address this issue, two candidate files were identified for modification.
Based on the issue description and the candidate files, the Manager defined two file-level tasks.
For these tasks, two Developers were assigned: Django Database Specialist (Developer I) and Alex Rossini (Developer II).
Following a kick-off meeting attended by both Developers and Managers, the Django Database Specialist commenced work first, followed by Alex Rossini.
During the coding phase, Developer I identified the code lines to be modified and generated the new code to replace them. The initial code changes made by Developer I were approved by the QA Engineer.
Developer II made three attempts during the coding process. The QA Engineer provided review comments on the first two attempts.
Ultimately, both Developers completed their coding tasks, and the merged results from their code changes passed all necessary tests.

\begin{figure}
        \centering
        \includegraphics[width=0.75\linewidth]{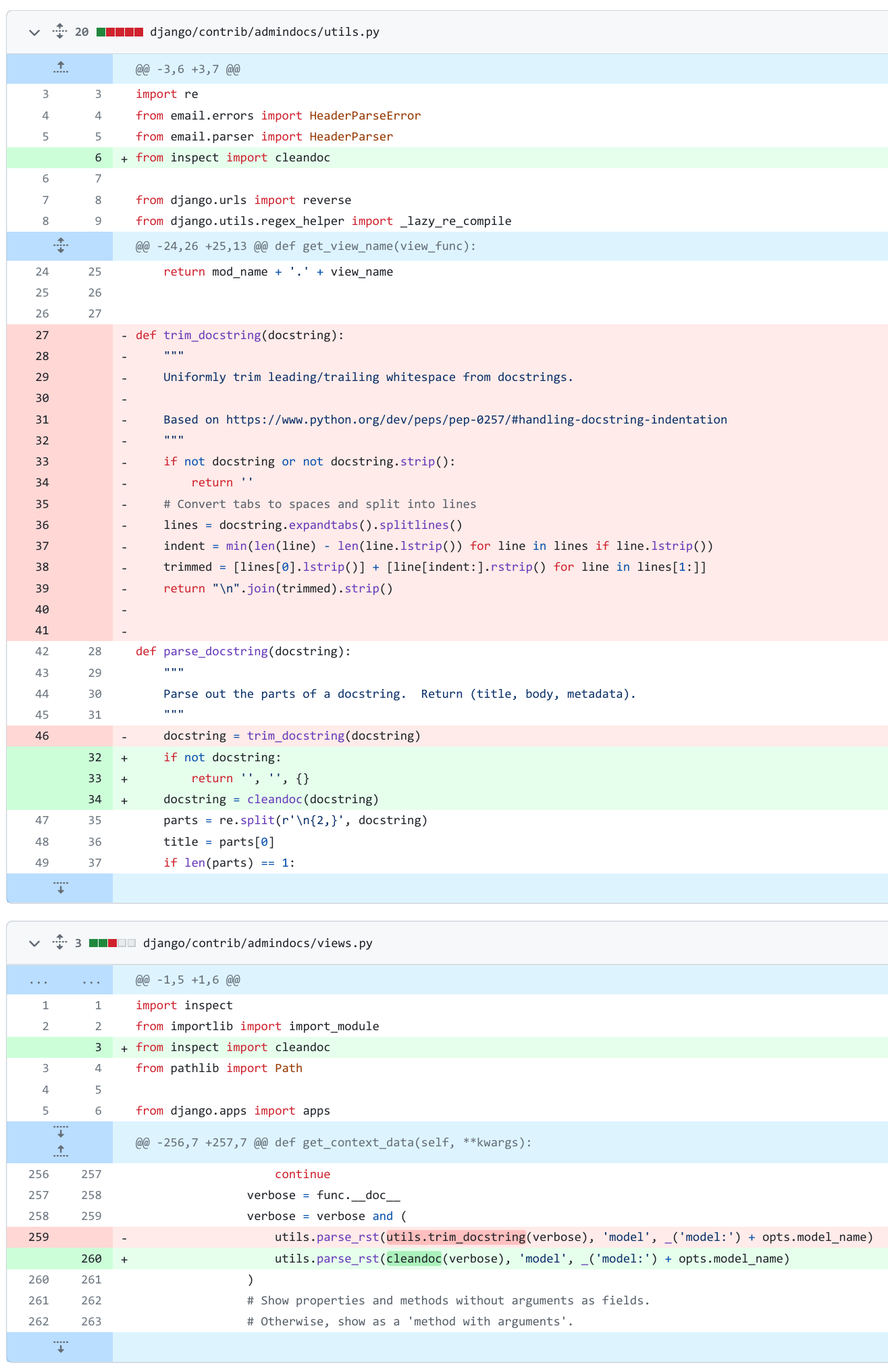}
        \caption{\small Case from \texttt{Django} (gold)~\citep{django12155}.}
        \label{fig:django_case_gold}
\end{figure}

\begin{figure}
        \centering
        \includegraphics[width=0.75\linewidth]{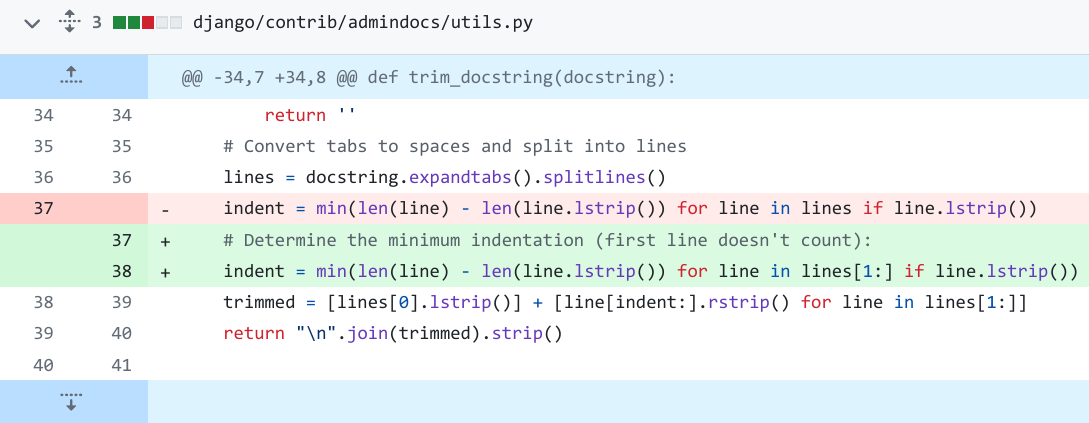}
        \caption{\small Case from \texttt{Django} (ours) for issue~\citep{django30255}.}
        \label{fig:django_case_our}
\end{figure}

\Fig~\ref{fig:django_case_gold} shows a reference issue resolution result, which resolves the issue~\footnote{\url{https://code.djangoproject.com/ticket/30255}} from the repository \texttt{Django}~\citep{django}, the human developer modifies four hunks in two files~\citep{django12155}. 
Despite the presence of modifications in two files, our method focuses on changes in only one file, as shown in Figure \ref{fig:django_case_our}. Notably, this simpler modification allows the repository to pass all necessary test cases.

\section{The performance of the QA Engineer Agent}\label{appendix:qa_case}

\Fig~\ref{fig:sklearn_case_gold} shows an issue~\citep{sklearn9784} from the repository \texttt{scikit-learn}~\citep{sklearn} and the reference code change~\citep{sklearn9288}. During the flow of our framework, the Developer firstly modifies the code as shown in \Fig~\ref{fig:sklearn_case_our_before_review} but the parameter\texttt{random\_state} (Line $371$ in the new-version code) of the function \texttt{kmeans\_single} is not assigned the right number in \texttt{seeds}. 
After the erroneous modification was made, the QA Engineer identified the mistake and provided feedback. Their commentary highlighted the issue: ``This code change modifies the implementation of K-means algorithm and doesn't seem entirely correct''. They further elaborated, ``Running the algorithm just one time could lead to worse results, compared to running it multiple times (n\_init times) and choosing the best result, as was originally done''. This critique specifically targets the flaw associated with the iterative process (``running times''). 
With the help of the QA Engineer, the Developer further revise the code, and the final code change is shown in \Fig~\ref{fig:sklearn_case_our_after_review}. All of the necessary test cases are passed after applying this code change.

\section{Related Work (Detailed)}\label{appendix:related}

\subsection{Large Language Models}
Large Language Models (LLMs) refer to the pre-trained language models that contain a large number of parameters~\citep{DBLP:journals/corr/abs-2303-18223}. The parameter counts of these models typically range in the tens or hundreds of billions. 
Popular LLMs include the Generative Pre-trained Transformer (GPT) series, such as GPT-3~\citep{radford2018improving}, GPT-4~\citep{openai2023gpt4}, and the open-source LLaMA~\citep{touvron2023llama} which publicly shares its weight information. The first version of the open-source model LLaMA has parameters ranging from 7 billion to 65 billion. Many researchers~\citep{llama-moe-2023, openlm2023openllama} have built upon the foundation of LLaMA, implementing enhancements to forge new LLMs.
These LLMs have demonstrated formidable natural language generation capabilities in general scenarios, with GPT-4, in particular, standing out~\citep{DBLP:conf/nips/LiuXW023, DBLP:conf/nips/ZhengC00WZL0LXZ23}. It has consistently maintained the top position in several rankings, including code generation, reflecting its significant potential in tasks related to software engineering~\citep{DBLP:journals/corr/abs-2308-10620}.

\subsection{LLM-Based Multi-Agent System}

With the powerful text generation capabilities of LLMs, many researchers~\citep{hong2023metagpt,DBLP:journals/corr/abs-2306-03314,DBLP:journals/corr/abs-2308-07201,DBLP:journals/corr/abs-2308-08155, qian2023communicative, tufano2024autodev, zhang2024unifying} have explored the construction of LLM-based Multi-Agent Systems, enabling them to accomplish tasks beyond the capabilities of the LLMs themselves. For example, MetaGPT~\citep{hong2023metagpt}, which simulates the Standardized Operating Procedures (SOPs) of a programming team, completing tasks including definition, design, planning, coding, and testing through constructed roles (e.g., product managers, architects, project managers, etc.). This framework has achieved leading scores on the HumanEval~\citep{HumanEval} and MBPP~\citep{MBPP}, outperforming many LLMs, and researchers show its ability to complete a software establishment (e.g., a code repository to play Gomoku game), indicating that a multi-agent framework can better leverage the capabilities of LLMs in code generation tasks. 
Moreover, \citet{qian2023communicative} designed ChatDev, a virtual development company simulating a human development team, which decomposes requirements into atomic tasks assigned to the developer agents. Developers mitigate the hallucination that may arise with the LLM through mutual communication and self-reflection mechanisms. Experimental results show that ChatDev can complete the establishment of some small projects (averaging no more than $5$ files per project) in a relatively short time (less than $7$ minutes on average). However, these works focus on the transformation from the requirements to code and overlook the code change generation during software evolution, which requires not only understanding the requirement but also dealing with the large repository.

\subsection{Automatic Bug Fixing}

\github issue resolution is a fundamental aspect of software evolution, with bug fixing being one of the most common scenarios. Fixing bugs involves both bug localization and repair. Previous researchers~\citep{DBLP:conf/icse/ZhouZL12, DBLP:journals/tr/QiSYZM22} have developed methods to localize bugs before modifying the code. DreamLoc, proposed by \citet{DBLP:journals/tr/QiSYZM22}, effectively models the characteristics of bug reports and source code files.
For automatic program repair, \citet{DBLP:conf/sigsoft/WongSKG21} explored a retrieval-based method, while \citet{DBLP:conf/icse/YeM24} proposed ITER, a generation-based method for handling fault localization re-execution. Additionally, some researchers~\citep{DBLP:conf/sigsoft/Wang0JH23, 0034WJH21} have combined retrieval techniques with generation models.
Recently, \citet{DBLP:conf/icse/XiaWZ23} demonstrated that directly applying popular LLMs significantly outperforms existing APR methods, showcasing their potential for generating diverse and effective patches. \citet{DBLP:journals/corr/abs-2403-17134} introduced RepairAgent, an autonomous LLM-based agent that plans and executes bug fixes by dynamically interacting with various tools.

\section{Limitation}\label{appendix:limitation}

\paragraph{Prompt} The design of prompt words may impact the performance of LLMs, thereby affecting the validity and fairness of the results~\citep{ChenCHC23}. While this paper focuses on innovative aspects of the proposed framework design and relies on practical guidelines for the design of prompt word templates~\citep{shieh2023best} to reduce the emergence of design biases, the complete elimination of the prompt bias is extremely difficult due to the inherent biases in the dataset instances and the limitations of API resources.

\paragraph{Dataset} The dataset contains a limited variety of software types. The evaluating dataset, \Swebench, encompasses $12$ repositories, which cover the Python programming language. However, this quantity remains insufficient compared to the diverse software projects available on \github. The code style, architectural design, and implementation techniques of these selected repositories, while representative, cannot fully reflect the diversity of all code repositories. In particular, the current dataset may fail to encompass some specialized fields or different programming paradigms, such as microservice architecture~\citep{DBLP:journals/tse/ZhouPXSJLD21} and functional programming~\citep{DBLP:conf/fpca/Johnsson87}. This limitation implies that, although our framework is designed to be independent of any specific software, the validation of its effectiveness and general applicability might be affected by this limited sample scope. Therefore, applying the findings of this paper to other code repositories may require further validation.